\documentclass[article]{jss}

\usepackage{orcidlink,thumbpdf,lmodern}
\usepackage{algorithm}
\usepackage{algpseudocode}
\usepackage{amsmath}
\usepackage{amsfonts}
\usepackage{enumitem}
\usepackage{makecell}
\usepackage{multirow}
\usepackage{comment}

\usepackage{framed}

\newcommand{\class}[1]{`\code{#1}'}
\newcommand{\fct}[1]{\code{#1()}}
\newcommand{\M}{\mathcal{M}}
\newcommand{\btheta}{\boldsymbol{\theta}}
\newcommand{\btau}{\boldsymbol{\tau}}
\DeclareMathOperator*{\argmax}{arg\,max}
\DeclareMathOperator*{\argmin}{arg\,min}

\newtheorem{remark}{Remark}
\newtheorem{definition}{Definition}

\author{Xingchi Li~\orcidlink{0009-0006-2493-0853} \\ Texas A\&M University
   \And Xianyang Zhang \\Texas A\&M University}
\Plainauthor{Xingchi Li and Xianyang Zhang}

\title{\pkg{fastcpd}: Fast Change Point Detection in \proglang{R}}
\Plaintitle{fastcpd: Fast Change Point Detection in R}
\Shorttitle{Fast Change Point Detection in \proglang{R}}

\Abstract{
  Change point analysis is concerned with detecting and locating structure
  breaks in the underlying model of a sequence of observations ordered by time,
  space or other variables. A widely adopted approach for change point analysis
  is to minimize an objective function with a penalty term on the number of
  change points. This framework includes several well-established procedures,
  such as the penalized log-likelihood using the (modified) Bayesian information
  criterion (BIC) or the minimum description length (MDL). The resulting
  optimization problem can be solved in polynomial time by dynamic programming
  or its improved version, such as the Pruned Exact Linear Time (PELT) algorithm
  \citep{killick2012optimal}. However, existing computational methods often
  suffer from two primary limitations: (1) methods based on direct
  implementation of dynamic programming or PELT are often time-consuming for
  long data sequences due to repeated computation of the cost value over
  different segments of the data sequence; (2) state-of-the-art \proglang{R}
  packages do not provide enough flexibility for users to handle different
  change point settings and models. In this work, we present the \pkg{fastcpd}
  package, aiming to provide an efficient and versatile framework for change
  point detection in several commonly encountered settings. The core of our
  algorithm is built upon PELT and the sequential gradient descent method
  recently proposed by \cite{zhang2023sequential}. We illustrate the usage of
  the \pkg{fastcpd} package through several examples, including mean/variance
  changes in a (multivariate) Gaussian sequence, parameter changes in
  regression models, structural breaks in ARMA/GARCH/VAR models,
  and changes in user-specified models.
}

\Plainkeywords{Change point analysis, Gradient descent, Quasi-Newton's method,
Regression models, Segmentation, Time series models}

\Address{
  Xingchi Li\\
  Department of Statistics\\
  Texas A\&M University\\
  College Station, Texas, USA\\
  E-mail: \email{anthony.li@stat.tamu.edu}\\
  URL: \url{https://xingchi.li/}\\

  Xianyang Zhang\\
  Department of Statistics\\
  Texas A\&M University\\
  College Station, Texas, USA\\
  E-mail: \email{zhangxiany@stat.tamu.edu}\\
  URL: \url{https://zhangxiany-tamu.github.io/}
}

\begin{document}


\section{Introduction}
\label{sec:intro}

Change point analysis is a classical but still very active research field. It
aims to detect and locate structural breaks in the underlying model of a
sequence of observations ordered by time, space, or other variables.
This field originated in the 1950s for industrial quality control and has since
become a critical tool across various fields such as signal processing, climate
science, economics, finance, medicine, and bioinformatics. The widespread
applicability of change point analysis is also evident from the adoption of
different terminologies, such as switch point problems
\citep{stephens1994bayesian, lindelov2020mcp}, broken-line relationships
\citep{muggeo2008segmented}, and anomaly detection \citep{shipmon2017time},
when researchers approach change point problems from various perspectives.
For readers who wish to gain more insights into the state-of-the-art
developments and recent innovations in change point detection problems,
they may refer to comprehensive book-length treatments and reviews, such as
\cite{brodsky1993nonparametric, csorgo1997limit, tartakovsky2014sequential, %
aue2013structural, niu2016multiple, aminikhanghahi2017survey, %
truong2020selective, liu2022high}.

Change point detection methods can be broadly classified into two categories:
online methods and offline methods. Online methods aim to detect changes as
soon as they have occurred in a real-time setting and are often used for event
or anomaly detection. On the other hand, offline methods are used to
retrospectively detect changes when all samples are received and are sometimes
referred to as signal segmentation. Here, we focus on the offline change point
detection methods, which often involve three major elements:
\begin{itemize}
  \item a cost function or a test statistic for quantifying the homogeneity of
    a segment;
  \item a penalty/constraint on controlling the number of change points;
  \item a search method for determining the optimal change point number
    and their locations.
\end{itemize}
Each of the above elements affects the overall performance of a
change point detection method \citep{truong2020selective}. A widely adopted
approach translates the offline change point detection (or signal segmentation)
problem into a model selection problem by solving a penalized optimization with
a penalty term on the number of change points. This framework includes several
well-established procedures, such as the penalized log-likelihood using the
(modified) Bayesian information criterion (BIC) or the minimum description
length (MDL).
Dynamic programming \citep{auger1989algorithms, jackson2005algorithm} offers
exact solutions to the penalized optimization problem when the penalty term is
linear in the number of change points. However, dynamic programming is often
computationally intensive. Pruning strategies like PELT
\citep{killick2012optimal, rigaill2010pruned} have been introduced to alleviate
the computational burden. Yet, they still face significant computational costs,
especially when (i) the data sequence is long and (ii) obtaining the cost value
involves solving a non-trivial optimization problem. In the worst-case scenario,
the computational cost of dynamic programming coupled with pruning
strategies remains in the same order $O(\sum_{t = 1}^T \sum_{s = 1}^t q(s))$ as
that without pruning, where $T$ denotes the length of the data sequence and
$q(s)$ represents the time complexity for calculating the cost function value
based on $s$ data points. Recently, \cite{zhang2023sequential} proposed a
sequential updating method by integrating gradient descent and quasi-Newton's
method with dynamic programming. The core idea is to update the cost value using
the information from previous steps, eliminating the need to re-optimize the
objective function over each data segment. This approach boosts computational
efficiency by reducing the time complexity to
$O(q_0\sum_{t = 1}^T \lvert R_t \rvert)$, where $q_0$ is the time complexity
for performing a one-step update described in Section~\ref{sec:algorithm} and
$|R_t|$ denotes the size of the set of candidate change point locations after
pruning at the $t$th step.

Existing \proglang{R} packages for change point analysis can be categorized
based on their detection algorithms, as discussed in \cite{bai2021multiple}.
One line of approaches involves finding the posterior probabilities of change
point locations, which can be achieved using packages such as \pkg{bcp}
\citep{erdman2008bcp, barry1993bayesian},
\pkg{mcp} \citep{lindelov2020mcp} and
\pkg{Rbeast} \citep{zhao2019detecting}. Another line of approaches involves
optimizing penalized cost functions and packages such as \pkg{changepoint}
\citep{killick2014changepoint},
\pkg{CptNonPar} \citep{mcgonigle2023nonparametric},
\pkg{fpop} \citep{maidstone2017optimal},
\pkg{gfpop} \citep{runge2023gfpop},
\pkg{strucchange}
\citep{zeileis2002strucchange} and \pkg{VARDetect} \citep{bai2021multiple},
fall under this category. Finally, some packages use multiscale analysis to
subsample and aggregate results, including
\pkg{breakfast} \citep{Fryzlewicz2018tguh},
\pkg{InspectChangepoint} \citep{wang2018high}
\pkg{not} \citep{baranowski2019narrowest}, \pkg{stepR} \citep{stepr} and
\pkg{wbs} \citep{Fryzlewicz2014wbs}.
Table~\ref{tab:package comparison} provides a detailed comparison of selected
offline change point analysis packages, including their functionalities,
speed, and accuracy.

\begin{table}[t!]
  \centering
  {
    \small
    \bgroup
    \def\arraystretch{0.9}
    \begin{tabular}{>{\hspace{-2pt}}c<{\hspace{-2pt}}|cc|ccc|c<{\hspace{-2pt}}>{\hspace{-2pt}}c<{\hspace{-2pt}}>{\hspace{-2pt}}c<{\hspace{-2pt}}>{\hspace{-2pt}}c|>{\hspace{-2pt}}c<{\hspace{-2pt}}>{\hspace{-2pt}}c<{\hspace{-2pt}}}
      \hline
      \multirow{2}{*}{\proglang{R} package}    & \multicolumn{2}{c|}{{\footnotesize basic characteristics}} & \multicolumn{3}{c|}{{\footnotesize regression models}}                           & \multicolumn{4}{c|}{{\footnotesize time series models}}                                            & \multirow{2}{*}{{\footnotesize speed}} & \multirow{2}{*}{{\footnotesize accuracy}} \\
                                               & {\footnotesize mean}   & {\footnotesize variance}   & {\footnotesize \code{lm}} & {\footnotesize \code{glm}} & {\footnotesize lasso} & {\footnotesize ar}    & {\footnotesize arma}  & {\footnotesize garch} & {\footnotesize var} &                                        &                                           \\ \hline
      {\footnotesize \pkg{fastcpd}           } & \checkmark             & \checkmark                 & \checkmark                & \checkmark                 & \checkmark            & \checkmark            & \checkmark            & \checkmark            & \checkmark          & $\star \star \star$                    & $\star \star \star$                       \\
      {\footnotesize \pkg{CptNonPar}         } & \checkmark             &                            &                           &                            &                       & \checkmark            & \checkmark            & \checkmark            &                     & $\star \star$                          & $\star \star \star$                       \\
      {\footnotesize \pkg{strucchange}       } & \checkmark             &                            & \checkmark                &                            & \checkmark            &                       &                       &                       &                     & $\star \star$                          & $\star \star$                             \\
      {\footnotesize \pkg{VARDetect}         } &                        &                            &                           &                            &                       &                       &                       &                       & \checkmark          & $\star \star \star$                    & $\star \star \star$                       \\
      {\footnotesize \pkg{Rbeast}            } & \checkmark             & \checkmark                 &                           &                            &                       &                       &                       &                       &                     & $\star \star \star$                    & $\star \star$                             \\
      {\footnotesize \pkg{ecp}               } & \checkmark             & \checkmark                 &                           &                            &                       &                       &                       &                       &                     & $\star$                                & $\star \star \star$                       \\
      {\footnotesize \pkg{gfpop}             } & 1                      & 1                          &                           &                            &                       &                       &                       &                       &                     & $\star \star \star$                    & $\star \star \star$                       \\
      {\footnotesize \pkg{mcp}               } & 1                      & 1                          &                           &                            &                       & \checkmark            &                       &                       &                     & $\star$                                & $\star \star \star$                       \\
      {\footnotesize \pkg{not}               } & 1                      & 1                          &                           &                            &                       &                       &                       &                       &                     & $\star \star \star$                    & $\star \star \star$                       \\
      {\footnotesize \pkg{changepoint}       } & 1                      & 1                          &                           &                            &                       &                       &                       &                       &                     & $\star \star \star$                    & $\star \star$                             \\
      {\footnotesize \pkg{bcp}               } & \checkmark             &                            &                           &                            &                       &                       &                       &                       &                     & $\star \star \star$                    & $\star \star \star$                       \\
      {\footnotesize \pkg{InspectChangepoint}} & \checkmark             &                            &                           &                            &                       &                       &                       &                       &                     & $\star \star \star$                    & $\star \star \star$                       \\
      {\footnotesize \pkg{jointseg}          } & \checkmark             &                            &                           &                            &                       &                       &                       &                       &                     & $\star \star \star$                    & $\star \star \star$                       \\
      {\footnotesize \pkg{breakfast}         } & 1                      &                            &                           &                            &                       &                       &                       &                       &                     & $\star \star$                          & $\star \star \star$                       \\
      {\footnotesize \pkg{wbs}               } & 1                      &                            &                           &                            &                       &                       &                       &                       &                     & $\star \star \star$                    & $\star \star \star$                       \\
      {\footnotesize \pkg{mosum}             } & 1                      &                            &                           &                            &                       &                       &                       &                       &                     & $\star \star \star$                    & $\star \star \star$                       \\
      {\footnotesize \pkg{fpop}              } & 1                      &                            &                           &                            &                       &                       &                       &                       &                     & $\star \star \star$                    & $\star \star \star$                       \\
      {\footnotesize \pkg{stepR}             } & 1                      &                            &                           &                            &                       &                       &                       &                       &                     & $\star \star \star$                    & $\star \star \star$                       \\
      {\footnotesize \pkg{segmented}         } &                        &                            &                           &                            &                       & \checkmark            &                       &                       &                     & $\star \star \star$                    & $\star$                                   \\ \hline
    \end{tabular}
    \egroup
  }
  \caption{Selective comparison of existing \proglang{R} packages for offline
    change point detection. \checkmark: supported; 1: only univariate data is
    supported; the number of stars indicates the speed and accuracy of the
    package, ranging from one (worst) to three (best) stars. We consider three
    settings:
    (i) mean and/or variance changes; (ii) changes in the parameters of linear
    models, generalized linear models (including logistic regression and Poisson
    regression), and penalized linear regression; (iii) changes in the structure
    of time series models including AR($p$), ARMA($p$, $q$), GARCH($p$, $q$) and
    VAR($p$) models. The comparisons are based on the simulation settings in the
    \pkg{fastcpd} vignettes\protect\footnotemark and the Well-log data set
    detailed in Section~\ref{subsec:well-log data}.
  }
  \label{tab:package comparison}
\end{table}

\footnotetext{\url{%
https://cran.r-project.org/package=fastcpd/vignettes/comparison-packages.html}}

Although the packages mentioned above excel in certain aspects, a package that
is both computationally efficient and capable of dealing with
a wide range of settings is still lacking. In this work, we develop
the \proglang{R} package \pkg{fastcpd} (\textbf{fast} \textbf{c}hange
\textbf{p}oint \textbf{d}etection) as a general framework for efficient change
point detection.

The \pkg{fastcpd} package offers a suite of built-in models for
change point detection under different settings, including
(i) mean and/or variance changes for multivariate data;
(ii) changes in the parameters of linear
models,
generalized linear models
(including logistic regression and Poisson regression),
and penalized linear regression; (iii) changes in time series models, including
ARMA (Autoregressive Moving Average), GARCH (Generalized Autoregressive
Conditional Heteroskedasticity), and VAR (Vector Autoregressive) models.
The \pkg{fastcpd} package allows custom cost functions, significantly enhancing
its adaptability to different real-world scenarios. Key features of the
package include the sequential updating method
for reduced time complexity, a general framework applicable to diverse data
types/models, and a user-friendly interface.

In Section~\ref{sec:algorithm}, we formulate the change point detection problem
as a model selection problem by optimizing a penalized objective function.
We present dynamic programming for solving the optimization problem and two
modifications: (i) the pruning strategy and (ii) the sequential gradient descent
to improve its computational efficiency. We discuss different choices for the
penalty terms, including the Bayesian information criterion (BIC),
the modified BIC, and the minimum description length (MDL).
Section~\ref{sec:models and software} describes the main function \fct{fastcpd}
in the package. We illustrate the use of our package in mean/variance change
models in Section~\ref{sec:mean variance change model}. In subsequent sections,
we detail the application of our package to various statistical models.
These include linear regression (Section~\ref{sec:changes-in-linear-models})
and generalized linear regression
(Section~\ref{sec:changes-in-generalized-linear-regression-models}).
Sections~\ref{sec:autoregressive model}-\ref{sec:vector autoregressive model}
illustrate the usage of the package for several time series models, including
AR, ARMA, GARCH, and VAR models.
In Section~\ref{sec:custom cost function}, we demonstrate the usage of the
package with a user-specified cost function. We analyze several real data
sets in Section~\ref{sec: real data analysis}. Section~\ref{sec:advanced usages}
explores several advanced usage scenarios, which include interpolating the
vanilla PELT with the sequential updating algorithm, adaptive epoch numbers,
and line search for sequential quasi-Newton's method.
Finally, the \pkg{fastcpd} package is available for download from CRAN
at \url{https://cran.r-project.org/package=fastcpd}. A cheat sheet
and comprehensive documentation can be found at
\url{https://fastcpd.xingchi.li}.

\section{Algorithm} \label{sec:algorithm}

Consider a sequence of $T$ observations $z_1, z_2, \ldots, z_T$ ordered in time.
We aim to partition the sequence into homogeneous segments such that
observations within each piece share the same (probabilistic) behavior.
One way to tackle the change point detection problem is to translate it into a
model selection problem. We denote by
$z_{s:t}=(z_s,z_{s+1},\dots,z_{t})$ the segment of data points observed from
time $s$ to time $t$ with $1\leq s\leq t\leq T$. Let $C(z_{s:t})$ be a cost
function for measuring the homogeneity of the segment $z_{s:t}$. Often, the
value of the cost function can be computed by solving the following
minimization problem
\begin{align}\label{eq:definitionofcostfunction}
C(z_{s:t}) = \min_{\theta \in \Theta} \sum_{i = s}^t l(z_i, \theta),
\end{align}
where $l(\cdot, \theta)$ is the individual loss function parameterized
by $\theta$ that belongs to a compact parameter space
$\Theta \subset \mathbb{R}^d$. A particular choice of $l(\cdot,\theta)$
is the negative log-likelihood of $z_i$.
When $z_i=(x_i,y_i)$ contains a set of predictors $x_i$ and a response $y_i$,
we can choose $l(z_i,\theta)=L(f(x_i,\theta),y_i)$, where $L$ is a loss function
and $f(\cdot,\theta)$ is an unknown regression function parameterized by
$\theta$ for predicting the response.

Suppose there are $k$ change points dividing the data sequence into $k + 1$
segments.
The cost value for the whole data sequence is defined as
\[
  C(k, T) =
  \min_{\boldsymbol{\tau}} \sum_{j = 0}^k C(z_{\tau_j + 1:\tau_{j + 1}}),
\]
where the minimization is over all possible change point locations
$\boldsymbol{\tau}=(\tau_1, \ldots, \tau_k)$ with
$1=\tau_0<\tau_1<\cdots<\tau_k<\tau_{k+1}=T$.
To estimate the number of change points, we consider the
following optimization problem
\[\min_{k} \{C(k,T) + f(k)\},\]
where $f(k)$ is a penalty term on the number of change points.

\textbf{Dynamic programming.} When $f(k)=\beta(k+1)$, the above penalized
optimization problem can be solved through dynamic programming
\citep{killick2012optimal,jackson2005algorithm}. To describe the procedure,
let us consider the data sequence up to time $t$, i.e., $z_{1:t}$. Denote
by $F(t)=\min_k \left\{C(k,t)+f(k)\right\}$ the minimum value of the
penalized cost for segmenting $z_{1:t}$. We derive a recursion for $F(t)$ by
conditioning on the last change point location,
\begin{align}
F(t):=&\min_{k,\boldsymbol{\tau}}\sum^{k}_{j=0}
       \left\{C(z_{\tau_j+1:\tau_{j+1}})+\beta\right\} \nonumber \\
=&\min_{k,\boldsymbol{\tau}}\left[\sum^{k-1}_{j=0}
    \left\{C(z_{\tau_j+1:\tau_{j+1}})+\beta\right\}+
      C(z_{\tau_k+1:t})+\beta\right]\nonumber
\\=&\min_{0\leq \tau\leq t-1}
     \left[\min_{\tilde{k},\boldsymbol{\tau}}\sum^{\tilde{k}}_{j=0}
     \left\{C(z_{\tau_j+1:\tau_{j+1}})+\beta\right\}+
        C(z_{\tau+1:t})+\beta\right] \nonumber
\\=&\min_{0\leq \tau\leq t-1}\left\{F(\tau)+C(z_{\tau+1:t})+\beta\right\},
   \label{eq1}
\end{align}
where $\tau_{k+1}=t$ in the first equation and $\tau_{\tilde{k}+1}=\tau$ in the
third equation.
The segmentations can be recovered by taking the argument $\tau$, which
minimizes (\ref{eq1}), i.e.,
\begin{align}\label{eq-taustar}
\tau^*=\argmin_{0\leq \tau\leq t-1}\left\{F(\tau)+C(z_{\tau+1:t})+\beta\right\},
\end{align}
which gives the optimal location of the last change point in segmenting
$z_{1:t}$. Let $\mathcal{C}(t)=\{\mathcal{C}(\tau^*),\tau^*\}$ be the estimated
change point locations associated with $z_{1:t}$. The procedure is repeated
until all the change point locations are identified. Our final estimates of the
change point locations for the whole data sequence
are given by $\mathcal{C}(T)$.

\textbf{Pruning.}
A popular way to improve the efficiency of dynamic programming is by pruning
the candidate set to find the last change point in each iteration. Suppose
for any $\tau<t<t'$, there exists a constant $c_0$ such that
\begin{align}\label{eq-check1}
C(z_{\tau+1:t})+C(z_{t+1:t'}) + c_0\leq C(z_{\tau+1:t'}).
\end{align}
\cite{killick2012optimal} showed that for some $t>\tau$ if
\begin{align}\label{eq-check2}
F(\tau)+C(z_{\tau+1:t}) + c_0 >F(t),
\end{align}
then at any future time point $t'>t$, $\tau$ can never be the optimal location
of the most recent change point prior to $t'$. To appreciate this, we note that
(\ref{eq-check1}) and (\ref{eq-check2}) together imply that
\begin{align*}
F(\tau) + C(z_{\tau+1:t'}) \geq
  F(\tau) +  C(z_{\tau+1:t})+C(z_{t+1:t'}) + c_0 > F(t) + C(z_{t+1:t'}),
\end{align*}
which suggests that $\tau$ is a sub-optimal choice as compared to $t$ in
segmenting $z_{1:t'}$. Motivated by this observation, one can define a
sequence of sets $\{R_t\}^{T}_{t=1}$ recursively as
\begin{align}
  R_t=\left\{\tau\in R_{t-1}: F(\tau)+C(z_{\tau+1:t-1}) + c_0\leq F(t-1)\right\}
    \cup\{t-1\}. \label{pruning}
\end{align}
Then $F(t)$ can be computed as
\begin{align}\label{eq}
F(t)=\min_{\tau\in R_{t}}\left\{F(\tau)+C(z_{\tau+1:t})+\beta\right\}
\end{align}
and the minimizer $\tau^*$ in (\ref{eq-taustar}) belongs to $R_t$.
The constant $c_0$ used in the pruning step can be selected as $c_0=0$ if the
cost function is defined as in (\ref{eq:definitionofcostfunction}).
More generally, consider
\[C(z_{s:t}) = \min_{\theta\in\Theta}
  \left\lbrace \sum_{i = s}^t l(z_i, \theta) + g_{s:t}(\theta) \right\rbrace\]
where $g_{s:t}(\theta)$ is a function of $\theta$ that can vary with respect to
the segment. An example for $g_{s:t}(\theta)$ is given in
Section~\ref{sec:penalized regression model} for sparse regression. Suppose
\begin{align*}
\max_{\theta\in\Theta}
  \left\{g_{\tau+1:t}(\theta) + g_{t+1:t'}(\theta)-g_{\tau+1:t'}(\theta)\right\}
  \leq -c_0,
\end{align*}
for all $\tau<t<t'$. Then we have
\begin{align*}
    &C(z_{\tau+1:t}) + C(z_{t+1:t'}) + c_0
    \\ \le& \min_{\theta\in\Theta} \left\lbrace \sum_{i = \tau+1}^{t'}
      l(z_i,\theta) + g_{\tau+1:t}(\theta) + g_{t+1:t'}(\theta) \right\rbrace +
        c_0 \\
    \le& \min_{\theta\in\Theta} \left\lbrace \sum_{i = \tau+1}^{t'}
      l(z_i,\theta) + g_{\tau+1:t'}(\theta) \right\rbrace +
        \max_{\theta\in\Theta}\left\{g_{\tau+1:t}(\theta) +
          g_{t+1:t'}(\theta) -g_{\tau+1:t'}(\theta)\right\} + c_0
    \\
    \leq&  C(z_{\tau+1:t'}).
\end{align*}

This pruning technique forms the basis of the Pruned Exact Linear Time (PELT)
algorithm. Under suitable conditions that allow the expected
number of change points to increase linearly with $T$, \cite{killick2012optimal}
proved that the expected computational cost for PELT is bounded by
$LT$ for some constant $L<\infty$.
In the worst case, where no pruning occurs, the computational cost of PELT is
the same as vanilla dynamic programming.

\textbf{Sequential gradient descent.} For large-scale data, the computational
cost of PELT can still be prohibitive due to the burden of repeatedly solving
the optimization problem (\ref{eq}). For many statistical models, the time
complexity for obtaining $C(z_{s:t})$ is linear in the number of observations
$t-s+1$. Therefore, in the worst-case scenario, the overall time complexity
can be as high as $O(T^3)$.
To alleviate the problem, \cite{zhang2023sequential} recently proposed a fast
algorithm by sequentially updating the cost function using a gradient-type
method to reduce the computational cost while maintaining a similar estimation
accuracy. Instead of repeatedly solving the optimization problem to obtain the
cost value for each data segment, \cite{zhang2023sequential}'s approach
updates the cost value using the parameter estimates from the
previous intervals.

For completeness, we derive the sequential gradient descent (SeGD) algorithm
here based on a heuristic argument. For $\tau\leq t-2$, suppose we have
calculated an approximate minimizer $\hat{\theta}_{\tau+1:t-1}$ for
$$\tilde{\theta}_{\tau+1:t-1}=\argmin_{\theta\in\Theta}%
  \sum^{t-1}_{i=\tau+1}l(z_i,\theta).$$
We want to find the cost value for the next data segment $z_{\tau+1:t}$,
\begin{align}\label{eq-3}
C(z_{\tau+1:t})=&\min_{\theta\in\Theta}\sum^{t}_{i=\tau+1}l(z_i,\theta)=\sum^{t}_{i=\tau+1}l(z_i,\tilde{\theta}_{\tau+1:t}),
\end{align}
where $\tau\geq 0$ and $t\leq T.$ Assume that $l(z,\theta)$ is twice differentiable in $\theta$ for any given $z$.
As $\tilde{\theta}_{\tau+1:t}$ is the minimizer of (\ref{eq-3}), it satisfies the first order condition (FOC)
$\sum^{t}_{i=\tau+1}\nabla l(z_i,\tilde{\theta}_{\tau+1:t})=0.$
Taking a Taylor expansion around $\hat{\theta}_{\tau+1:t-1}$ in the FOC, we obtain
\begin{align*}
0=&\sum^{t}_{i=\tau+1}\nabla l(z_i,\tilde{\theta}_{\tau+1:t}) \\
\approx
&\sum^{t}_{i=\tau+1}\nabla l(z_i,\hat{\theta}_{\tau+1:t-1})+\sum^{t-1}_{i=\tau+1}\nabla^2 l(z_i,\hat{\theta}_{\tau+1:t-1})(\tilde{\theta}_{\tau+1:t}-\hat{\theta}_{\tau+1:t-1}) \\
\approx &\nabla l(z_t,\hat{\theta}_{\tau+1:t-1})+\sum^{t-1}_{i=\tau+1}\nabla^2 l(z_i,\hat{\theta}_{\tau+1:t-1})(\tilde{\theta}_{\tau+1:t}-\hat{\theta}_{\tau+1:t-1}),
\end{align*}
where $\sum^{t-1}_{i=\tau+1} \nabla l(z_i,\hat{\theta}_{\tau+1:t-1})\approx0$ as $\hat{\theta}_{\tau+1:t-1}$ is an approximate minimizer of $\sum^{t-1}_{i=\tau+1}l(z_i,\theta)$, and we drop the term $\nabla^2 l(z_t,\hat{\theta}_{\tau+1:t-1})$. Rearranging the terms, we get
\begin{equation}\label{eq-approx1}
\begin{split}
&\tilde{\theta}_{\tau+1:t}\approx \hat{\theta}_{\tau+1:t-1}
-\left(\sum^{t-1}_{i=\tau+1}\nabla^2 l(z_i,\hat{\theta}_{\tau+1:t-1})\right)^{-1}\nabla l(z_t,\hat{\theta}_{\tau+1:t-1}).
\end{split}
\end{equation}
As the RHS of (\ref{eq-approx1}) does not necessarily fall into the parameter space $\Theta$, we suggest a projection step. Specifically, let $\mathcal{P}_{\Theta}(\theta)$ denote the projection of any $\theta\in\mathbb{R}^d$ onto $\Theta$.
The above observation motivates us to consider the following update
\begin{align}
\hat{\theta}_{\tau+1:t}=\mathcal{P}_{\Theta}(\hat{\theta}_{\tau+1:t-1}-H_{\tau+1:t-1}^{-1}\nabla l(z_t,\hat{\theta}_{\tau+1:t-1})),  \label{projectedupdatestep}
\end{align}
where $H_{\tau+1:t-1}$ is a preconditioning matrix that serves as a surrogate for the second-order information
$\sum^{t-1}_{i=\tau+1}\nabla^2 l(z_i,\hat{\theta}_{\tau+1:t-1})$. When the second-order information is available, we suggest updating the preconditioning matrix through the iteration
\begin{align*}
H_{\tau+1:t} = H_{\tau+1:t-1}+  \nabla^2 l(z_{t},\hat{\theta}_{\tau+1:t}).
\end{align*}
Alternatively, using the idea of Fisher scoring, one can also update the preconditioning matrix through
\begin{align*}
H_{\tau+1:t} =H_{\tau+1:t-1}+ \mathcal{I}_{t}(\hat{\theta}_{\tau+1:t}),
\end{align*}
where $\mathcal{I}_{t}(\theta)=E[\nabla^2 l(z_{t},\theta)|x_t]$ with $x_t$ being a subvector of $z_t$ such as the covariates in the regression setting. Finally, we approximate $\tilde{\theta}_{\tau+1:t}$ by $(t-\tau)^{-1}\sum^{t}_{j=\tau+1}\hat{\theta}_{\tau+1:j}$ and the cost value $C(z_{\tau+1:t})$ by
\begin{align*}
\widehat{C}(z_{\tau+1:t})=\sum^{t}_{i=\tau+1}l\left(z_i,(t-\tau)^{-1}\sum^{t}_{j=\tau+1}\hat{\theta}_{\tau+1:j}\right).
\end{align*}

In the above derivation, a single pass of each data point is performed during the update of the cost value, leading to low data utilization and inaccurate approximation when the length of the data segment is short. It is natural to pass the data multiple times to increase the data utilization and the approximation accuracy at the cost of higher time complexity. Algorithm \ref{alg:fastcpd} summarizes the details of the algorithm with multiple passes/epochs. Specifically, we will use each data point $W + 1 \geq 1$ times in updating the parameter estimates for a particular segment. Here, the superscript $(w, t)$ denotes the corresponding value in the $w$-th epoch at time step $t$, and the superscript $(W)$ represents the corresponding value after the $W$-th pass. We can also allow an adaptive number of multiple passes for data segments with different lengths; see Section~\ref{subsec:advanced usage adaptive number of epochs} for the details.

\begin{algorithm}
  \caption{Sequential Updating Algorithm with Multiple Epochs}
  \label{alg:fastcpd}
  \begin{algorithmic}
  \Require Data $\{z_i\}_{i=1}^T$, individual loss function $l(\cdot, \theta)$,
    the penalty constant $\beta$ and the number of epochs $W$.
  \Ensure Locations of the change points.
  \State Initialize $F=(-\beta, 0)$, $\mathrm{cp}=\{\emptyset, \{0\}\}$, and $R=\{\{0\}, \{0, 1\}\}$.
  \For{$t\gets 2,\ldots,T$}
    \For{$\tau\in R_t \setminus \{t - 1\}$}
        \State Initialize $\hat{\theta}_{\tau+1:\tau+1}^{(0,t)}$
    and
      $H_{\tau+1:\tau+1}^{(0,t)}$.
      \For{$w\gets 0,\ldots,W$}
        \For{$j\gets \tau + 2,\ldots, t$}
          \State $\hat{\theta}_{\tau + 1:j}^{(w,t)} = \mathcal{P}_{\Theta}%
            (\hat{\theta}_{\tau + 1:j - 1}^{(w,t)} - %
            H_{\tau + 1:j-1}^{(w,t),-1}\nabla %
            l(z_j, \hat{\theta}_{\tau + 1:j-1}^{(w, t)}))$,
          \State $H_{\tau + 1:j}^{(w,t)} = H_{\tau + 1:j - 1}^{(w,t)} + %
                  \nabla^2 l(z_j, \hat{\theta}_{\tau + 1:j}^{(w,t)})$,
        \State for $w=W$, $S_{\tau + 1:j}^{(W)} = S_{\tau + 1:j-1}^{(W)} + %
        \hat{\theta}_{\tau + 1:j}^{(W,t)}$ with $S_{\tau + 1:\tau+1}^{(W)}=\hat{\theta}_{\tau + 1 : \tau+1}^{(W, t)}$.
        \EndFor
            \State For $w\leq W-1$, $(\hat{\theta}_{\tau + 1 : \tau+1}^{(w+1, t)}, %
            H_{\tau + 1:\tau+1}^{(w+1, t)})=(\hat{\theta}_{\tau + 1:t}^{(w, t)}, %
            H_{\tau + 1:t}^{(w, t)})$.
      \EndFor
    \EndFor
    \For{$\tau \in R_t$}
      \State $\hat{C}(z_{\tau + 1:t})=\sum_{i = \tau + 1}^t l(z_i, %
        (t - \tau)^{-1} S_{\tau + 1:t}^{(W)})$.
    \EndFor
    \State $F(t) = \min_{\tau \in R_t}\{ F(\tau) + %
      \hat{C}(z_{\tau + 1:t}) + \beta\}$.
    \State $\tau^\ast = \argmin_{\tau \in R_t}\{F(\tau) + %
      \hat{C}(z_{\tau + 1:t}) + \beta\}$.
    \State $\mathrm{cp}(t) = \{\mathrm{cp}(\tau^\ast), \tau^\ast\}$.
    \State $R_{t + 1} = \{\tau \in R_t : F(\tau) + %
      C(z_{\tau + 1:t}) + c_0 \le F(t)\}\cup \{t\}$.
  \EndFor
  \State \Return $\mathrm{cp}(T)$
  \end{algorithmic}
\end{algorithm}

\subsection{Penalty term}\label{sec:penalty-selection}
In this section, we discuss a few well-adopted choices of penalty functions in the literature. To be clear, we will focus on the case where $C(z_{s:t})=\min_{\theta\in\Theta}\sum^{t}_{i=s}l(z_i,\theta)$ with $l(z_i,\theta)$ being the negative log-likelihood function of $z_i$.
Recall that we aim to solve the following $l_0$ penalized optimization problem
\begin{align*}
\min_k\min_{\boldsymbol{\tau}}\left\{\sum^{k}_{j=0}\min_{\theta_j\in\Theta}\sum^{\tau_{j+1}}_{i=\tau_j+1}l(z_i,\theta_j) + f(k)\right\}
\end{align*}
where we simultaneously optimize over the number of change points $k$, the locations of the change points $\boldsymbol{\tau}$, and the parameters within each segment. With $k$ change points that divide the data sequence into $k+1$ segments, the total number of parameters is equal to $(k+1)d+k$, where $(k+1)d$ counts the number of parameters from the $k+1$ segments. The traditional BIC criterion sets $f(k)=\{(k+1)d+k\}\log(T)/2$ or equivalently (to up an additive constant term) $f(k)=\beta(k+1)$
\begin{align}
    \beta=(d+1)\log(T)/2. \label{eq:betavalueinbic}
\end{align}

\cite{zhang2007modified} derived a modified BIC criterion (mBIC) for selecting the number of change points in a Gaussian location model with piece-wise constant mean values. In a similar spirit, we can extend the mBIC to general likelihood models.

\begin{definition}[Modified BIC]\label{def:mbic}
{\rm
Given the number of change points $k$ and the change point locations $\boldsymbol{\tau}$, the mBIC is defined as
    \begin{align}
        \mathrm{mBIC}(k,\boldsymbol{\tau}):=& \sum^{k}_{j=0}C(z_{\tau_{j}+1:\tau_{j+1}}) + \frac{d}{2}\sum^{k}_{j=0}\log (\tau_{j+1}-\tau_j) + (k+1)\log(T). \label{definition1eq}
    \end{align}
}
\end{definition}
In Appendix~\ref{sec:mbic for mean change with fixed variance}, we provide a heuristic derivation for the mBIC and show that it coincides with the mBIC proposed in \cite{zhang2007modified} for the Gaussian location model considered therein.

\begin{remark}
{\rm
Suppose the dimension of the parameter in the $j$th segment is $d_j$, which is allowed to vary over segments. In this case, the mBIC can be defined in the same way with the term $(d/2)\sum^{K}_{j=0}\log (\tau_{j+1}-\tau_j)$ being replaced by $\sum^{K}_{j=0}(d_j/2)\log (\tau_{j+1}-\tau_j)$.
}
\end{remark}

Another commonly used criterion for selecting the number of change points is the so-called minimum description length (MDL), which selects the best model with the shortest description of the data, see, e.g., \cite{davis2006structural}. In the current context, the MDL can be defined in the following way.
\begin{definition}[MDL]\label{def:mdl}
{\rm
Given the number of change points $k$ and the change point locations $\boldsymbol{\tau}$, the MDL is defined as
    \begin{align*}
        \mathrm{MDL}(k,\boldsymbol{\tau}):=& \sum^{k}_{j=0}C(z_{\tau_{j}+1:\tau_{j+1}}) + \frac{d}{2}\sum^{k}_{j=0}\log_2(\tau_{j+1}-\tau_j)
        + (k+1)\{\log_2(T)+\log_2(d)\}
        \\&+ \log_2(k).
    \end{align*}
}
\end{definition}
Suppose $d$ and $k$ are bounded (i.e., they do not grow with $T$). Ignoring the smaller order terms, the MDL is approximately equal to
\begin{align*}
&\sum^{k}_{j=0}C(z_{\tau_{j}+1:\tau_{j+1}}) + \frac{d}{2}\sum^{k}_{j=0}\log_2(\tau_{j+1}-\tau_j) + (k+1)\log_2(T)
\\=&\sum^{k}_{j=0}C(z_{\tau_{j}+1:\tau_{j+1}}) + \frac{d}{2\log(2)}\sum^{k}_{j=0}\log(\tau_{j+1}-\tau_j) + \frac{k+1}{\log(2)}\log(T).
\end{align*}

Dynamic programming and its variants introduced in the previous section can still be used to solve the optimization problems associated with mBIC and MDL. More precisely, we define the adjusted cost functions and the penalty coefficients as
\begin{align*}
    \text{mBIC: }& \widetilde{C}(z_{\tau_{j}+1:\tau_{j+1}})=C(z_{\tau_{j}+1:\tau_{j+1}})+ \frac{d}{2}\log(r_{j+1}-r_j),& \tilde{\beta} &= (d+2)\log(T)/2; \\
    \text{MDL: }& \breve{C}(z_{\tau_{j}+1:\tau_{j+1}})=C(z_{\tau_{j}+1:\tau_{j+1}})+\frac{d}{2}\log_2(r_{j+1}-r_j), &\breve{\beta}&=(d+2)\log_2(T)/2,
\end{align*}
for the data segment $z_{{\tau_{j}+1:\tau_{j+1}}}$, where $r_j=\tau_j/T$.
To implement dynamic programming, one only has to replace the original cost function and the penalty coefficient with either $\widetilde{C}$ and $\tilde{\beta}$ (for mBIC) or $\breve{C}$ and $\breve{\beta}$ (for MDL). Suppose $C$ satisfies \eqref{eq-check1}. In the case of mBIC, setting $\tilde{c}_0 = c_0 + d\log(2)$, we have
\begin{align*}
\widetilde{C}(z_{\tau+1:t})+\widetilde{C}(z_{t+1:t'}) + \tilde{c}_0 &= C(z_{\tau+1:t}) + C(z_{t+1:t'}) + c_0 + \frac{d}{2}\log\left(\frac{4(t-\tau)(t'-t)}{T^2}\right) \\
&\le C(z_{\tau+1:t'}) + \frac{d}{2}\log((t'-\tau)/T) + \frac{d}{2}\log\left(\frac{4(t-\tau)(t'-t)}{T(t'-\tau)}\right) \\
&\le \widetilde{C}(z_{\tau+1:t'}).
\end{align*}
where we have used the fact that $(t-\tau)(t'-t)\leq (t'-\tau)^2/4$. Thus, we consider the following update:
\begin{align*}
F(t) = \min_{\tau \in \widetilde{R}_t} \{F(\tau)+\widetilde{C}(z_{\tau+1:t})+\tilde{\beta}\},
\end{align*}
where $\widetilde{R}_t= \{\tau \in \widetilde{R}_{t - 1} : F(\tau) + \widetilde{C}(z_{\tau + 1:t - 1}) + \tilde{c}_0 \le F(t-1)\}\cup \{t - 1\}$.

Tabel~\ref{tab:penaltycomparison} summarizes the three penalty criteria with the cost function being a negative log-likelihood function, where the ``cost adjustment'' is defined as $\widetilde{C}(\cdot) - C(\cdot)$ and $\breve{C}(\cdot) - C(\cdot)$ for mBIC and MDL, respectively.
Note that $r_{j + 1} - r_j = (\tau_{j + 1} - \tau_j) / T\leq 1$, leading to a negative adjustment term in both mBIC and MDL. All three criteria are implemented in the \pkg{fastcpd} package.

\begin{table}[t!]
\centering
\begin{tabular}{cccc}
\hline
Penalty & Cost Adjustment   & $\beta_T$ & Pruning Adjustment             \\ \hline
BIC     &      0             & $(d + 1) \log(T) / 2$ & 0  \\
mBIC    & $d \log(r_{j+1}-r_j)/2$   & $(d + 2) \log(T) / 2$  & $d\log(2)$ \\
MDL     & $d \log_2(r_{j+1}-r_j)/2$ & $(d + 2) \log_2(T) / 2$ & $d\log_2(2)$ \\ \hline
\end{tabular}
\caption{\label{tab:penaltycomparison} A comparison of the three penalty criteria, where the ``cost adjustment'' is given by $\widetilde{C}(\cdot) - C(\cdot)$ and $\breve{C}(\cdot) - C(\cdot)$ and ``pruning adjustment'' is giving by $\tilde{c}_0 - c_0$ and $\breve{c}_0 - c_0$ for mBIC and MDL respectively. The length of the whole data sequence is $T$ and
the number of parameters in each segment is denoted by $d$.
Here $r_{j+1} - r_j$ represents the length of the segment normalized by $T$. mBIC denotes the Modified BIC in Definition~\ref{def:mbic} and MDL denotes the minimum description length in Definition~\ref{def:mdl}.}
\end{table}

\section{Main function} \label{sec:models and software}

\pkg{fastcpd} provides a versatile framework applicable to a broad spectrum of data types and modeling scenarios. Specifically, it offers
the built-in functionality to perform change point analysis in several scenarios, including detecting changes in the (multivariate) mean and/or variance, linear regression, penalized linear regression, logistic regression, Poisson regression, ARIMA, GARCH, and VAR models. Each model has their corresponding function, as illustrated in the following sections. Table~\ref{tab:models} summarizes all the families users can use.

\begin{table}[t!]
\centering
{
  \small
  \begin{tabular}{ccp{7cm}}
    \hline
    Model & Family
      & Description \\ \hline
    \multirowcell{3}{Mean change\\Variance change\\Mean and variance change} & \multirowcell{3}{\code{mean}\\\code{variance}\\\code{mv/meanvariance}}
      & \code{formula = ~ .\ - 1} \\ && \multirow{2}{*}{\parbox{7cm}{These families are applicable to multivariate data.}} \\ &&\\ \hline
    \multirowcell{6}{Linear regression\\Logistic regression\\Poisson regression\\LASSO} & \multirowcell{6}{\code{lm}\\\code{binomial}\\\code{poisson}\\\code{lasso}}
      & \code{formula = y ~ .\ - 1} \\ && \multirow{5}{*}{\parbox{7cm}{These families require a labeled data set where the first column corresponds to the response variable, and the rest are covariates. \code{"binomial"} family accepts $y \in [0, 1]$ while \code{"poisson"} family requires $y \in \mathbb{N}$.}} \\ &&\\ &&\\ &&\\ &&\\ \hline
    \multirowcell{6}{AR($p$)\\ARMA($p$, $q$)\\ARIMA($p$, $d$, $q$)\\GARCH($p$, $q$)\\VAR($p$)} & \multirowcell{6}{\code{ar}\\\code{arma}\\\code{arima}\\\code{garch}\\\code{var}}
      & \code{formula = ~ .\ - 1} \\ && \multirow{5}{*}{\parbox{7cm}{These families are for time series data. The format of the input data can be a vector or a matrix for AR, ARIMA and GARCH families and a matrix (with columns corresponding to different time series) for the VAR family.}} \\ &&\\ &&\\ &&\\ &&\\\hline
    \multirowcell{3}{Custom model} & \multirowcell{3}{\code{custom}}
      & \multirow{3}{*}{\parbox{7cm}{\code{cost} must be specified to use a custom cost function. \code{cost\_gradient} and \code{cost\_hessian} are required to implement SeGD.}} \\ &&\\ &&\\\hline
  \end{tabular}
}
\caption{\label{tab:models} Models implemented in the \pkg{fastcpd} package. There are three categories of models based on different data types, namely (i) unlabeled data for mean and/or variance change models, (ii) labeled data for regression models, and (iii) time-ordered data for time series models.
Users can either use \fct{fastcpd} with the specified family or utilize
the corresponding \fct{fastcpd.family} functions.}
\end{table}

\fct{fastcpd} is the main function of the \pkg{fastcpd} package. The parameter list and explanations for each of the parameters are given below.

\begin{Code}
fastcpd(formula = y ~ . - 1, data, beta = "MBIC", cost_adjustment = "MBIC",
  family = NULL, cost = NULL, cost_gradient = NULL, cost_hessian = NULL,
  line_search = c(1), lower = rep(-Inf, p), upper = rep(Inf, p),
  pruning_coef = 0, segment_count = 10, trim = 0.02, momentum_coef = 0,
  multiple_epochs = function(x) 0, epsilon = 1e-10, order = c(0, 0, 0),
  p = ncol(data) - 1, cp_only = FALSE, vanilla_percentage = 0,
  warm_start = FALSE, ...)
\end{Code}

Not all the parameters can co-exist, and some parameters are used for a specific
family, which will be discussed in detail in the corresponding sections below.
Descriptions of a selective list of crucial parameters are provided here with
their suggested default values.
\begin{itemize}
  \item \code{formula}: A formula object specifying the model to be fitted. The
    (optional) response variable should be on the LHS of the formula, while the
    covariates should be on the RHS. The naming of variables used in the formula
    should be consistent with the column names in the data frame provided in
    \code{data}. The intercept term should be removed from the formula.
    The response variable is not needed for mean/variance change models and time
    series models. By default, an intercept column will be added to the data,
    similar to the \code{lm} function in \proglang{R}. Thus, it is suggested
    that users should remove the intercept term by appending \code{- 1} to the
    formula. Note that the \fct{fastcpd.family} functions do not require a
    formula input.
  \item \code{data}: A data frame of dimension $T \times d$ containing the data
  to be segmented (where each row denotes a data point $z_t \in \mathbb{R}^d$
  for $t = 1, \ldots, T$) is required in the main function, while a matrix or a
  vector input is also accepted in the \fct{fastcpd.family} functions.
  \item \code{beta}: Penalty criterion for the number of change points.
      This parameter takes a string value of
    \code{"BIC"}, \code{"MBIC"}, \code{"MDL"}
    or a numeric value. If a numeric value is provided, the value will be
    used as $\beta$
    in Algorithm~\ref{alg:fastcpd}. By default, the mBIC
    criterion is used, where $\beta=(d + 2) \log(T) / 2$.
    This parameter usage should be paired with \code{cost_adjustment} described
    below. Discussions about the penalty criterion can be found
    in Section~\ref{sec:penalty-selection}.
  \item \code{cost_adjustment}: Cost adjustment criterion.
    It can be \code{"BIC"}, \code{"MBIC"}, \code{"MDL"} or \code{NULL}.
    By default, the cost adjustment criterion is set to be \code{"MBIC"}.
    The \code{"MBIC"} and \code{"MDL"} criteria modify the cost function by
    adding a negative adjustment
    term to the cost function. \code{"BIC"} or \code{NULL} does not modify the
    cost function. Details can
    in found in Section~\ref{sec:penalty-selection}.
  \item \code{family}: Family class of the change point model. It can be
      \code{"mean"} for mean change,
      \code{"variance"} for variance change,
      \code{"meanvariance"} or \code{"mv"},
      for mean and/or variance change, \code{"lm"} for linear regression,
      \code{"binomial"} for logistic regression, \code{"poisson"} for
      Poisson regression, \code{"lasso"} for penalized linear regression,
      \code{"ar"} for AR($p$) models, \code{"ma"} for MA($q$) models,
      \code{"arma"} for
      ARMA($p$, $q$) models, \code{"arima"} for ARIMA($p$, $d$, $q$) models,
      \code{"garch"}
      for GARCH($p$, $q$) models, \code{"var"} for VAR($p$) models and
      \code{"custom"}
      for user-specified custom models.
      Omitting this parameter is the same as specifying the parameter to be
      \code{"custom"} or \code{NULL}, in which case, users must specify the
      custom cost function.
  \item \code{cost}: Cost function to be used. \code{cost},
  \code{cost_gradient}, and \code{cost_hessian} should not be specified at the
  same time with
  \code{family} as built-in families have cost functions implemented in C++
  to provide better performance. If not specified, the default is the negative
  log-likelihood for the corresponding family. Custom cost functions can be
  provided in the following two formats:
      \begin{itemize}
        \item \code{cost = function(data) \{...\}}
        \item \code{cost = function(data, theta) \{...\}}
      \end{itemize}
      Users can specify a loss function using the second format that will be
      used to calculate the cost value as in \eqref{eq-3}.
      In both formats, the input data is a subset of the original data frame in
      the form of a matrix (a matrix with a single column in
  the case of a univariate data set). In the first format, the specified cost
  function directly calculates the cost value. \fct{fastcpd} performs the
  vanilla PELT algorithm, and \code{cost_gradient} and \code{cost_hessian}
  should not be provided since no parameter updating is necessary for vanilla
  PELT. In the second format, the loss function $\sum^{t}_{i=s}l(z_i,\theta)$ is
  provided, which has to be optimized over the parameter $\theta$ to obtain the
  cost value. A detailed discussion about the custom cost function usage can be
  found in Section~\ref{sec:custom cost function}.
  \item \code{cost_gradient}: Gradient of the custom cost function.
    Example usage:
    \begin{Code}
    cost_gradient = function(data, theta) {
      ...
      return(gradient)
    }
    \end{Code}
  The gradient function takes two inputs, the first being a matrix representing
  a segment of the data, similar to the format used in the \code{cost} function,
  and the second being the parameter that needs to be optimized. The gradient
  function returns the value of the gradient of the loss function, i.e.,
  $\sum_{i = s}^t \nabla l(z_i, \theta)$.
    \item \code{cost_hessian}: Hessian of the custom loss function.
      The Hessian function takes two inputs, the first being a matrix
      representing a segment of the data, similar to the format used in the
      \code{cost} function, and the second being the parameter that needs to be
      optimized. The gradient function returns the Hessian of the loss function,
      i.e., $\sum_{i = s}^t \nabla^2 l(z_i, \theta)$.
  \item \code{line_search}: If a vector of numeric values is provided, a line
  search will be performed to find the optimal step size for each update.
  Detailed usage of \code{line_search} can be found in
  Section~\ref{subsec:advanced usage line search}.
  \item \code{lower}:
  Lower bound for the parameters. Used to specify the domain of the parameters
  after each gradient descent step. If not specified, the lower bound is set to
  be \code{-Inf} for all parameters. \code{lower} is especially useful when the
  estimated parameters take only positive values, such as the noise variance.
  \item \code{upper}: Upper bound for the parameters. Used to specify the domain
  of the parameters after each gradient descent step. If not specified, the
  upper bound is set to be \code{Inf} for all parameters.
  \item \code{pruning_coef}: Pruning coefficient $c_0$ used in the pruning step
  of the PELT algorithm in Eq~\eqref{pruning} with the default value 0.
  If \code{cost_adjustment} is specified as \code{"MBIC"}, an adjustment term
  $d\log(2)$ will be added to the pruning coefficient. If \code{cost_adjustment}
  is specified as
\code{"MDL"}, an adjustment term $d\log_2(2)$ will be added to the pruning
coefficient. Detailed discussion about the pruning coefficient can be found in
Section~\ref{sec:penalty-selection}.
  \item \code{segment_count}: An initial guess of the number of segments. If not
  specified, the initial guess of the number of segments is 10. The initial
  guess affects the initial estimates of the parameters in SeGD.
  \item \code{multiple_epochs}: A function can be specified such that an
  adaptive number of multiple epochs can be utilized to improve the algorithm's
  performance. \code{multiple_epochs} is a function of the length of the data
  segment. The function returns an integer indicating how many epochs should be
  performed apart from the default update (see Algorithm~\ref{alg:fastcpd}). By
  default, the function returns zero, meaning no multiple epochs will be used to
  update the parameters. Example usage:
    \begin{Code}
    multiple_epochs = function(segment_length) {
      if (segment_length < 100) 1
      else 0
    }
    \end{Code}
  This function will let SeGD perform parameter updates with an additional epoch
  for each segment with a length less than 100 and no additional epoch for
  segments with lengths greater or equal to 100.
  \item \code{vanilla_percentage}: The parameter $v$ is between zero and one.
  For each segment, when its length is no more than $vT$, the cost value will be
  computed by performing an exact minimization of the loss function over the
  parameter.
  When its length is greater than $vT$, the cost value is approximated through
  SeGD. Therefore, this parameter induces an algorithm that can be interpreted as
  an interpolation between dynamic programming with SeGD ($v=0$) and the vanilla
  PELT ($v=1$).
  The readers are referred to Section~\ref{subsec:advanced usage vanilla pelt}
  for more details.
  \item \code{warm_start}: If \code{TRUE}, the algorithm will use the estimated
parameters from the previous segment as the initial value for the
current segment. This parameter is only used for the \code{"glm"} families.
  \item \code{...}: Other parameters for specific models. One use case
is \code{include.mean}, which determines
if a mean/intercept term should be included in the ARIMA($p$, $d$, $q$)
or GARCH($p$, $q$) models.
\end{itemize}
The return value of the main function is a \class{fastcpd} object, which can be used to
perform further visualization and downstream analysis. The \class{fastcpd} object contains the following outputs: \code{call}, \code{data}, \code{family}, \code{cp_set},
\code{cost_values}, \code{residuals}, \code{thetas}, and \code{cp_only}. In particular,
\begin{itemize}
  \item \code{cp_set}: The set of change point locations.
  \item \code{cost_values}: The cost values for each segment.
  \item \code{residuals}: The residuals of the fitted model with the estimated change points. Only for the built-in families.
  \item \code{thetas}: The estimated parameters for each segment. Only for the built-in families.
\end{itemize}
Utility functions, including \fct{plot}, \fct{print} and \fct{summary} are provided as a part of the package, where a \class{fastcpd} object can be passed to these functions to produce a plot of the response values with the change point locations, a simple summary of the model and the estimated change points, or a detailed summary of the model including the estimated change points, the corresponding cost values, and the estimated parameters for each segment.

\section{Mean and variance change models}
\label{sec:mean variance change model}
Detecting mean change, variance change, and mean-variance change are fundamental scenarios in change point analysis. The \pkg{fastcpd} package provides built-in functions that can handle these settings with multivariate data.

We first discuss the multivariate mean change model with a constant but unknown covariance matrix.
We
illustrate the usage of the package through a simple example here.
Consider a data sequence with $T = 1000$ and three-dimensional observations generated from
three different multivariate Gaussian distributions with the mean vectors
$\boldsymbol{\mu}_1 = (0, 0, 0)$,
$\boldsymbol{\mu}_2 = (50, 50, 50)$ and $\boldsymbol{\mu}_3 = (2, 2, 2)$
and the covariance matrices
$\Sigma_1 = \Sigma_2 = \Sigma_3 = \mathrm{diag}(100, 100, 100)$, i.e.,
\begin{alignat*}{4}
  \boldsymbol{x}_t &\sim \mathcal{N}(\boldsymbol{\mu_1}, \Sigma_1) &
    \qquad & 1\leq t\leq 300, \\
  \boldsymbol{x}_t &\sim \mathcal{N}(\boldsymbol{\mu_2}, \Sigma_2) &
     & 301   \leq t \leq 700, \\
  \boldsymbol{x}_t &\sim \mathcal{N}(\boldsymbol{\mu_3}, \Sigma_3) &
     & 701\leq t \leq 1000.
\end{alignat*}
The two change points are located at $t = 300$ and $t = 700$.
The \fct{fastcpd.mean} function implements the vanilla PELT for detecting mean change, with the cost being the minimum negative log-likelihood of the multivariate Gaussian model. Specifically, for the data segment $\boldsymbol{x}_{s:t}$, the cost function is given by
\begin{align*}
    C(\boldsymbol{x}_{s:t}) &=  \frac{1}{2} \sum_{i = s}^t (\boldsymbol{x}_i - \bar{\boldsymbol{x}}_{s:t})^\top \hat{\Sigma}^{-1} (\boldsymbol{x}_i - \bar{\boldsymbol{x}}_{s:t})
    +\frac{(t - s + 1)d}{2}\log(2\pi) +\frac{t-s+1}{2}\log(|\hat{\Sigma}|),
\end{align*}
where $\bar{\boldsymbol{x}}_{s:t}=(t-s+1)^{-1}\sum^{t}_{i=s}\boldsymbol{x}_i$, $|\cdot|$ denotes the determinant of a matrix and
$\hat{\Sigma}$ is the Rice estimator of $\Sigma$
\begin{align*}
\hat{\Sigma}=\frac{1}{2(T-1)}\sum^{T-1}_{t=1}(\boldsymbol{x}_{t+1}-\boldsymbol{x}_t)(\boldsymbol{x}_{t+1}-\boldsymbol{x}_t)^\top.
\end{align*}
See Section~\ref{subsec: variance-estimation-with-change points} for a detailed discussion of the difference-based variance estimation and its extension to linear models with changes in the regression coefficients.
\begin{Schunk}
\begin{Sinput}
R> result <- fastcpd.mean(mean_data)
R> summary(result)
\end{Sinput}
\begin{Soutput}
Call:
fastcpd.mean(data = mean_data)

Change points:
300 700

Cost values:
2558.102 3417.845 2551.517
\end{Soutput}
\end{Schunk}

Next, we consider the variance change model, where the mean vector is fixed but unknown. Suppose we have a data sequence $\{\boldsymbol{x}_t\}_{t = 1}^T$ with $\boldsymbol{x}_t\in\mathbb{R}^d$. The cost function here for data segment $\boldsymbol{x}_{s:t}$ is given by
\begin{align*}
    C(\boldsymbol{x}_{s:t}) = \frac{t-s+1}{2} \left\{ d\log(2\pi) + d + \log \left(\left\lvert \frac{1}{t-s+1}\sum_{i = s}^t (\boldsymbol{x}_i-\bar{\boldsymbol{x}}_{1:T})(\boldsymbol{x}_i-\bar{\boldsymbol{x}}_{1:T})^\top \right\rvert\right)\right\}
\end{align*}
We generate a data sequence of length $T=1000$ from three-dimensional Gaussian distributions with zero mean and varying covariance matrices. Specifically,
\begin{alignat*}{4}
  \boldsymbol{x}_t &\sim \mathcal{N}(\boldsymbol{0}, \Sigma_1) &
    \qquad &1 \leq t \le 300, \\
  \boldsymbol{x}_t &\sim \mathcal{N}(\boldsymbol{0}, \Sigma_2) &
    & 301 \leq t\leq 700, \\
  \boldsymbol{x}_t &\sim \mathcal{N}(\boldsymbol{0}, \Sigma_3) &
    & 701 \leq t \leq 1000, \\
  \Sigma_i &= \tilde{\Sigma}_i^\top \tilde{\Sigma}_i &
    &\left(\tilde{\Sigma}_i\right)_{jk} \sim U[-1, 1] \qquad i = 1, 2, 3, \quad j, k = 1, \ldots, d,
\end{alignat*}
where $d = 3$. Again, the two change
points are located at $t = 300$ and $t = 700$. The \fct{fastcpd.variance} function implements the vanilla PELT for detecting variance change, with the cost being the minimum negative log-likelihood of the multivariate Gaussian model.
\begin{Schunk}
\begin{Sinput}
R> result <- fastcpd.variance(variance_data)
R> summary(result)
\end{Sinput}
\begin{Soutput}
Call:
fastcpd.variance(data = variance_data)

Change points:
300 700

Cost values:
753.6359 1706.144 1280.607
\end{Soutput}
\end{Schunk}
Finally, we consider a model where both the mean vector and the covariance matrix are allowed to change over time. The cost function for the data segment $x_{s:t}$ in this case is defined as
\begin{align*}
    C(\boldsymbol{x}_{s:t}) = \frac{t - s + 1}{2} \left\{ d \log(2\pi) + d + \log \left(\left\lvert \frac{1}{t-s+1}\sum_{i = s}^t (\boldsymbol{x}_i-\bar{\boldsymbol{x}}_{s:t})(\boldsymbol{x}_i-\bar{\boldsymbol{x}}_{s:t})^\top \right\rvert\right) \right\}.
\end{align*}
We generate a data sequence of length $T=2000$ from four-dimensional Gaussian distributions with time-varying mean and covariance matrix:
\begin{alignat*}{4}
  \boldsymbol{x}_t &\sim
    \mathcal{N}(0 \cdot \boldsymbol{1}_4, I_4)
    &\qquad 1\leq t &\le 300,\\
  \boldsymbol{x}_t &\sim
    \mathcal{N}(10 \cdot \boldsymbol{1}_4, I_4)
    &   301 \leq t & \leq 700, \\
  \boldsymbol{x}_t &\sim
    \mathcal{N}(0 \cdot \boldsymbol{1}_4, 100 \times I_4)
    &   701 \leq t & \leq 1000,  \\
  \boldsymbol{x}_t &\sim
    \mathcal{N}(0 \cdot \boldsymbol{1}_4, I_4)
    &   1001 \leq t & \leq 1300,  \\
  \boldsymbol{x}_t &\sim
    \mathcal{N}(10 \cdot \boldsymbol{1}_4, I_4)
    &  1301 \leq t & \leq 1700,  \\
  \boldsymbol{x}_t &\sim
    \mathcal{N}(10 \cdot \boldsymbol{1}_4, 100 \times I_4)
    &  1701 \leq t & \leq 2000.
\end{alignat*}
The change points are located at $300$, $700$, $1000$, $1300$, and $1700$, where each change point is associated with a specific
type of structural break, namely, mean change, variance change, and mean-variance change. The \fct{fastcpd.meanvariance} function implements the vanilla PELT for detecting mean-variance change, with the cost value being the minimum negative log-likelihood of the multivariate Gaussian model.
\begin{Schunk}
\begin{Sinput}
R> meanvariance_result <- fastcpd.meanvariance(mean_variance_data)
R> summary(meanvariance_result)
\end{Sinput}
\begin{Soutput}
Call:
fastcpd.meanvariance(data = mean_variance_data)

Change points:
300 700 1000 1300 1700

Cost values:
1714.968 2299.119 4500.027 1654.917 2259.023 4444.359
\end{Soutput}
\end{Schunk}
The \fct{fastcpd.meanvariance} correctly identifies all the change point locations. To detect those change points associated with only variance change,
one can use \fct{fastcpd.variance}, which successfully detect the variance change at $700$, $1000$ and $1700$.
\begin{Schunk}
\begin{Sinput}
R> variance_result <- fastcpd.variance(mean_variance_data)
R> summary(variance_result)
\end{Sinput}
\begin{Soutput}
Call:
fastcpd.variance(data = mean_variance_data)

Change points:
700 1000 1700

Cost values:
5645.66 4607.095 5546.242 4525.403
\end{Soutput}
\end{Schunk}

\section{Linear models} \label{sec:changes-in-linear-models}

Linear models are of fundamental importance in statistical analysis. In this section, we show how to use the \pkg{fastcpd} package
to detect changes in the coefficients of linear regression models. Specifically, we focus on the model
\begin{align}\label{lm}
y_t = x_t^\top \theta_t + \epsilon_t,\quad
\epsilon_t \sim \mathcal{N}(0, \sigma^2), \quad
1 \le t \le T,
\end{align}
where $x_t\in\mathbb{R}^d$ and the regression coefficients $\theta_t$s are assumed to be piece-wise constant over time $t$, i.e., there exists a set of change points $(\tau_{1}^*,\dots,\tau_{k*}^*)$ such that
$$\theta_{1}=\cdots=\theta_{\tau_1^*}\neq \theta_{\tau_1^*+1}=\cdots=\theta_{\tau_2^*}\neq \cdots \neq \theta_{\tau^*_{k^*}+1}=\cdots=\theta_T.$$
Denote by $z_t=(x_t,y_t)$ the observation at time $t$. Under the Gaussian assumption on the errors, we define the cost function for the segment $z_{s:t}$ to be the minimum value of the negative log-likelihood
\begin{align*}
C(z_{s:t})=\min_{\theta} \sum_{i = s}^t \left( \frac{1}{2}\log(2\pi\hat{\sigma}^2) + \frac{(y_i -  x_i^\top \theta)^2}{2\hat{\sigma}^2} \right)=\sum_{i = s}^t \left( \frac{1}{2}\log(2\pi\hat{\sigma}^2) + \frac{(y_i -  x_i^\top \hat{\theta})^2}{2\hat{\sigma}^2} \right),
\end{align*}
where $\hat{\theta}=(\sum^{t}_{i=s}x_ix_i^\top)^{-1}\sum^{t}_{i=s}x_iy_i$ is the least square estimator and $\hat{\sigma}^2$ is the Rice estimator of the error variance; see more details in the next subsection.

The \fct{fastcpd.lm} function is specifically designed for change point analysis in the above linear models. We illustrate its usage through a simple numerical example.
We generate observations from model (\ref{lm}) with $x_t\sim \mathcal{N}(0,1)$ (i.e., $p=1$), $\sigma^2=1$ and $T=300.$ The change points are at $t=100$ and $200$, and $\theta_t=1, -1, 0.5$ for the three segments respectively.
The following code shows how to use the \fct{fastcpd.lm} function to detect structural breaks in the regression coefficients.
\begin{Schunk}
\begin{Sinput}
R> result <- fastcpd.lm(cbind(y, x))
R> summary(result)
\end{Sinput}
\begin{Soutput}
Call:
fastcpd.lm(data = cbind(y, x))

Change points:
100 201

Cost values:
48.16996 66.1816 45.66268

Parameters:
  segment 1  segment 2 segment 3
1 0.9520606 -0.8307605 0.4593161
\end{Soutput}
\end{Schunk}
The \fct{summary} function applied to a \class{fastcpd} object returns the estimated change points, the estimated parameters for each segment, and the corresponding cost values. The \fct{plot} function applied to a \class{fastcpd} object produces a set of plots  of the estimated coefficients, responses, residuals, and covariates, together with
the estimated change point locations (marked with vertical grey lines).
The estimated change point locations divide the data into three segments. The residual plot is based on the linear regression models fitted on each segment. It is a diagnostic tool for checking the goodness of fit of the linear regression model with piecewise constant regression coefficients (a homogeneous residual plot suggests a good fit of the model to the data). When $p>1$, only the response values and residuals are plotted.
\begin{figure}
\centering
\begin{Schunk}
\begin{Sinput}
R> plot(result)
\end{Sinput}
\end{Schunk}
\includegraphics{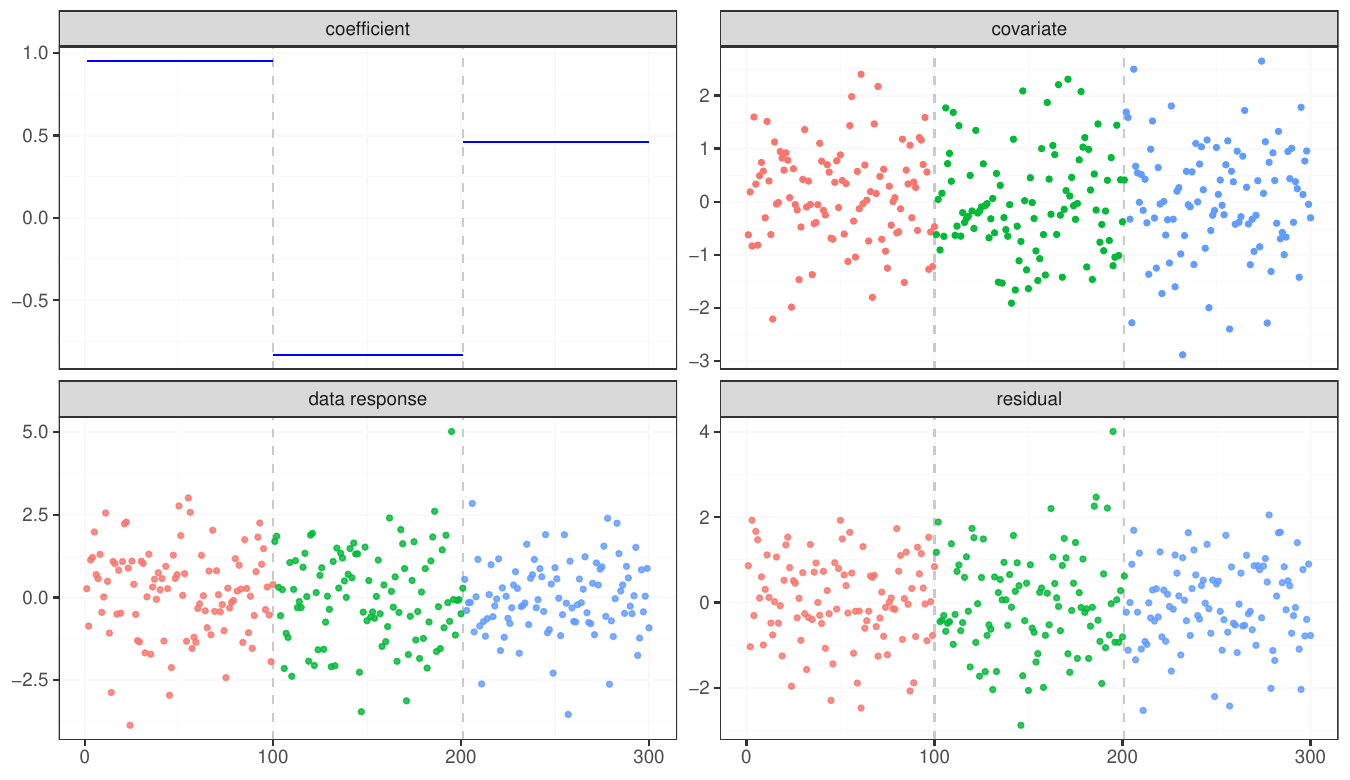}
\caption{Plots of the estimated coefficients, covariates, residuals, and responses together with the estimated change point locations (marked with vertical grey lines).
}
\label{fig:linear-regression-with-one-dimensional-covariate-example-plot}
\end{figure}

\subsection{Variance estimation}
\label{subsec: variance-estimation-with-change points}

In both the mean change models with constant variance and the linear models with changing coefficients but constant variance, the unknown variance must be estimated to compute the cost value.
Classical variance estimators are no longer reliable in the presence of change points.
In the nonparametric regression literature, there have been several proposals for difference-based variance estimation \citep{rice1984bandwidth, gasser1986residual, muller1987estimation, hall1990asymptotically, dai2014variance}, which are robust to potential structural break in the conditional mean function. As a simple illustration, consider the model
$$y_t=\mu_t+\epsilon_t,\quad 1\leq t\leq T,$$
where $\epsilon_i$s have zero mean and common variance $\sigma^2.$ \cite{rice1984bandwidth} proposed the first-order difference-based variance estimator
\begin{align}
  \hat{\sigma}^2_{\mathrm{Rice}} =
  \frac{1}{2(T - 1)} \sum_{t = 1}^{T - 1} (y_{t + 1} - y_t)^2, \label{riceestimator}
\end{align}
which has been shown to be consistent under change point models \citep{muller1999discontinuous}.

We extend the Rice estimator to the linear regression setup by considering the consecutive difference between two least squares estimators based on a sliding window containing $M$ data points.
In particular, we define
\begin{align*}
\hat{\theta}_{t,M}=\left(\sum^{t+M-1}_{i=t}x_ix_i^\top\right)^{-1}\sum^{t+M-1}_{i=t}x_iy_i,
\end{align*}
where $M$ is the window size such that $M\geq d.$ Define
\begin{align*}
    \hat{\sigma}^2_t & =\frac{
    (\hat{\theta}_{t+1,M}-\hat{\theta}_{t,M})^\top     (\hat{\theta}_{t+1,M}-\hat{\theta}_{t,M})
  }{
    \mathrm{tr}(H_{t+1,M} + H_{t,M} - 2 C_{t,M})
  },
\end{align*}
for $t=1,2,\dots,T-M$, where $H_{t,M}=(\sum^{t+M-1}_{i=t}x_ix_i^\top)^{-1}$ and
\begin{align*}
C_{t,M}=H_{t,M}\left(\sum^{t+M-1}_{i=t+1}x_ix_i^\top\right)H_{t+1,M}.
\end{align*}
We propose the following generalized Rice estimator
\begin{align*}
\hat{\sigma}^2_{\text{G-Rice}} &= \frac{1}{T - M} \sum_{t = 1}^{T - M} \hat{\sigma}^2_t.
\end{align*}

\begin{remark}
{\rm
We remark that the idea can be directly extended to multivariate multiple linear models, where the response is multi-dimensional.
}
\end{remark}

To illustrate the performance of this variance estimator, let us consider model (\ref{lm}) with
\begin{align}
  x_t \sim \mathcal{N}(0, I_d), \quad
  \epsilon_t \sim \mathcal{N}(0, 10^2), \quad
  1 \le t \le T,
\end{align}
for $d=3$ and $T=1000$. The true change
points are located at $300$ and $700$. True $\theta_t$'s for the three segments are defined respectively as
\begin{alignat*}{4}
  \theta_t &= (10, 1.2, -1) &1\le t&\le 300, \\
  \theta_t &= (-1, 8, 0.5) & 301 \le t &\le 700, \\
  \theta_t &= (0.5, -3, 0.2) \quad & 701 \le t &\le 1000.
\end{alignat*}
With $M=5$, the estimated variance is 103.0231, which is relatively close to the truth.
\begin{Schunk}
\begin{Sinput}
R> (variance_estimator <- variance.lm(cbind(y, x)))
\end{Sinput}
\begin{Soutput}
[1] 103.0231
\end{Soutput}
\end{Schunk}
We remark that this variance estimation method has been implemented in \fct{fastcpd} for linear models and \fct{fastcpd.lm} as a default.

\subsection{High-dimensional linear regression }
\label{sec:penalized regression model}

In this section, we illustrate how our package is useful in detecting change points in high-dimensional linear regression. More precisely, we consider the following model
\begin{align}\label{hd-lm}
y_t = x_t^\top \theta_t + \epsilon_t,\quad
1 \le t \le T,
\end{align}
where $\epsilon_t$ is mean zero with variance $\sigma^2$, $x_t\in\mathbb{R}^d$ and the regression coefficients $\theta_t$'s are assumed to be piece-wise constant over time $t$. We assume that $\theta_t$ is sparse and $d$ can be large as compared to $T$. The cost function for the data segment $z_{s:t}$ is defined as
\begin{align*}
    C(z_{s:t}) = \min_{\theta\in \Theta} \frac{1}{2} \sum_{i = s}^t (y_i - x_i^\top \theta )^2 + \lambda_{s:t} \sum_{i = 1}^d \lvert \theta_i \rvert,
\end{align*}
where $\theta=(\theta_1,\dots,\theta_d)$ and
$\lambda_{s:t} = \hat{\sigma} \sqrt{2 \log(d) / (t - s+1)}$ with $\hat{\sigma}$ being an estimate of the noise level. The constant $c_0$ in the pruning condition \eqref{eq-check1} can be specified as $$c_0=-1.3\hat{\sigma}\sqrt{2\log(d)}\max_{\theta\in\Theta} \sum_{i = 1}^d \lvert \theta_i \rvert,$$
where we have used the fact that $\sqrt{1/n_1}+\sqrt{1/n_2}-\sqrt{1/(n_1+n_2)}\leq 1.3$ for all integers $n_1,n_2\geq 1$.
In this case, we shall update the parameters using the sequential proximal gradient descent
introduced in Section 3.3 of \cite{zhang2023sequential}.

\begin{figure}
\centering
\begin{Schunk}
\begin{Sinput}
R> plot(result)
\end{Sinput}
\end{Schunk}
\includegraphics{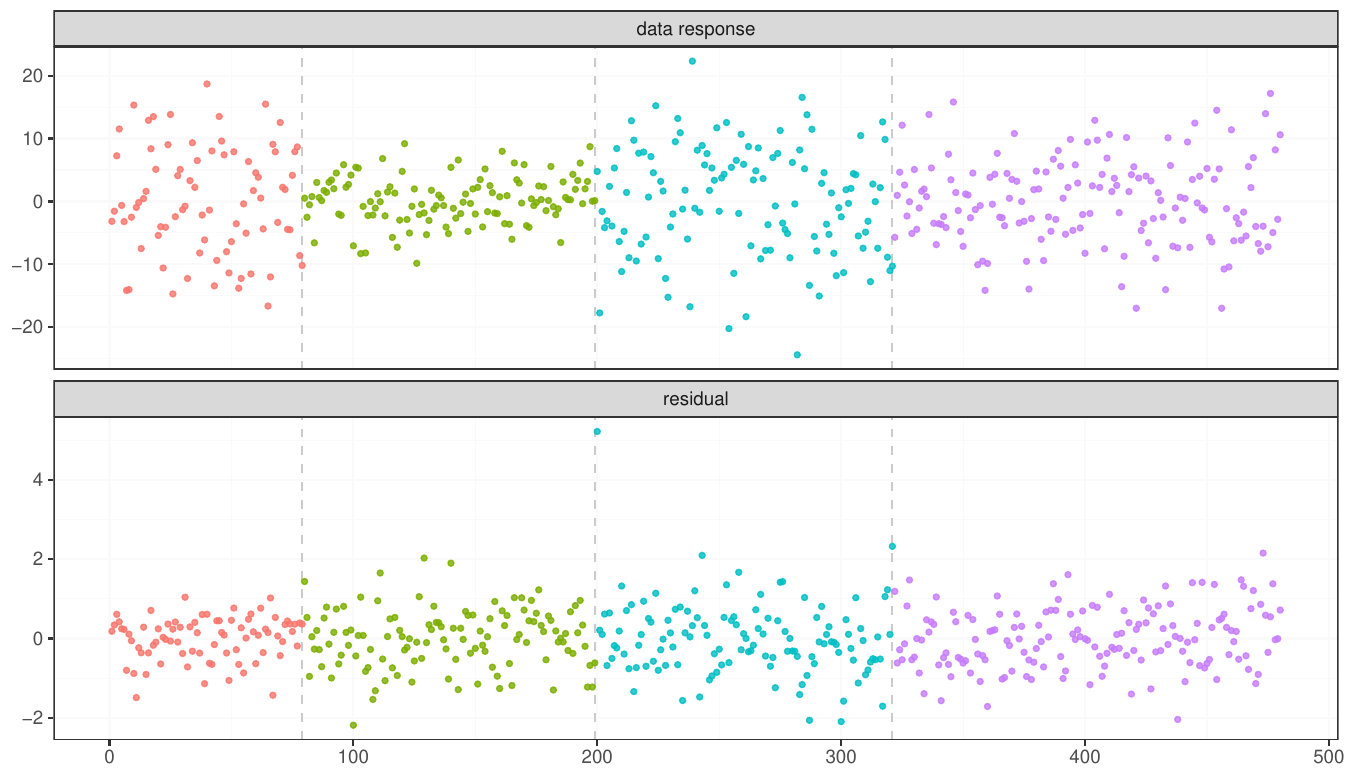}
\caption{Plots of the responses and residuals together with the estimated change point locations marked with vertical grey lines.}
\label{fig:penalized-regression-example-plot}
\end{figure}

\begin{figure}
\centering
\includegraphics{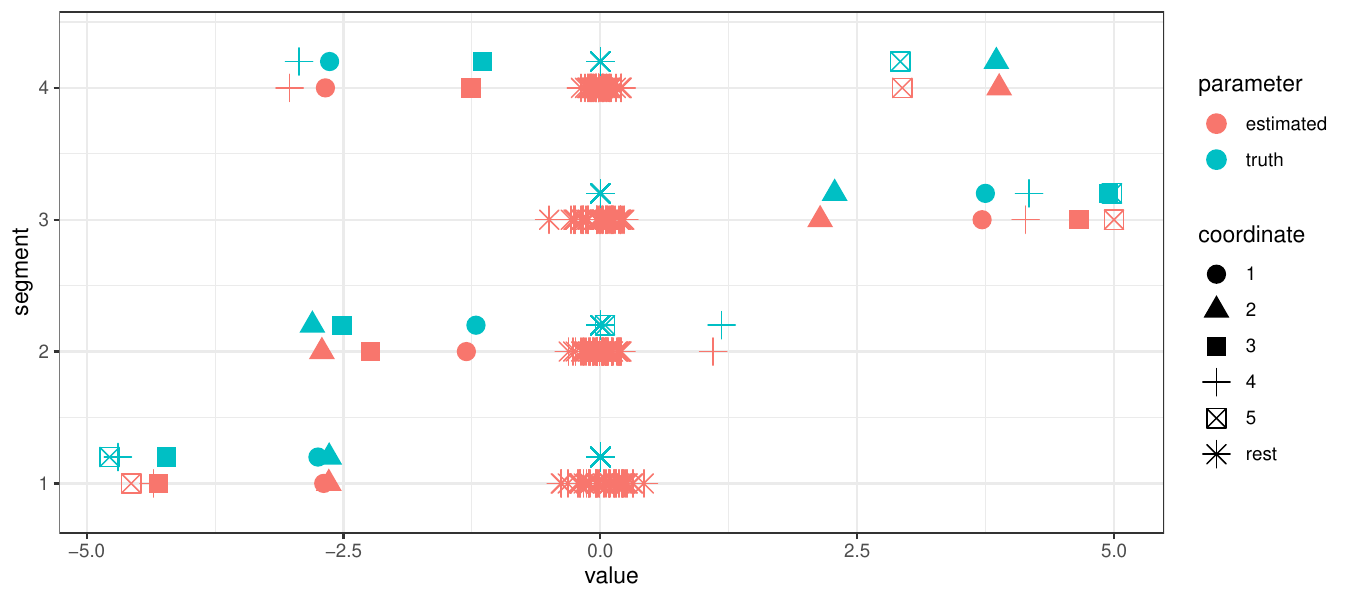}
\caption{Comparison between the estimated and true coefficients for each segment. The first five
coordinates are marked with distinct shapes, while the rest are marked with the same shape.}
\label{lassocoordinate}
\end{figure}

As an illustration, we generate data from model (\ref{hd-lm}) with $x_t\sim\mathcal{N}_p(0,I_d)$ and $\epsilon_t\sim\mathcal{N}(0,1)$ for $d=50$ and $T=480.$
The regression coefficient $\theta_t$ is piece-wise constant with the change points located at $80$, $200$, and $320$. Within each segment, the first five components of $\theta_t$ are nonzero, and
the remaining components are zero across all segments.
The following code illustrates the use of the \fct{fastcpd.lasso} function.
\begin{Schunk}
\begin{Sinput}
R> result <- fastcpd.lasso(cbind(y, x))
R> result@cp_set
\end{Sinput}
\begin{Soutput}
[1]  79 199 321
\end{Soutput}
\end{Schunk}
The \fct{plot} function (applied to the \class{fastcpd} object) returns the plots of the responses and residuals together with the estimated change point locations marked with vertical grey lines in Figure~\ref{fig:penalized-regression-example-plot}.
The residual plot suggests that the sparse linear model with the estimated change point locations provides a good fit to the data. Figure~\ref{lassocoordinate} shows the estimated coefficients for each segment and the true coefficients with each coordinate marked with different shapes.

\section{Generalized linear regression models}
\label{sec:changes-in-generalized-linear-regression-models}

In generalized linear models (GLM), the response variable $y_t$ is
assumed to follow a distribution from the canonical exponential family
\begin{align*}
  f(y_t; \gamma_t, \phi) = \exp
    \left\{\frac{y_t\gamma_t - b(\gamma_t)}{w^{-1} \phi} + c(y_t, \phi)\right\},
\end{align*}
where $\gamma_t$ is the canonical parameter, $\phi$ is the dispersion
parameter and $w$ is some known weight. The mean of $y_t$ can be obtained
through a known link function $g(\cdot)$, i.e.,
\begin{align*}
  g(E[y_t]) = g(\nabla b(\gamma_t)) = x_t^\top \theta_t,
\end{align*}
where the regression coefficients $\theta_t$s are assumed to be piece-wise constant over time $t$. Denote by $z_t=(x_t,y_t)$ for $1\leq t\leq T.$
When $\phi$ is known, the cost function is given by
\begin{align*}
    C(z_{s:t}) = \min_\theta - \sum_{i = s}^t \left\{ \frac{y_i\gamma_i - b(\gamma_i)}{w^{-1} \phi} + c(y_i, \phi) \right\},\quad g(\nabla b(\gamma_i)) = x_i^\top \theta.
\end{align*}
When $\phi$ is unknown and can be estimated by $\hat{\phi}$ from the data, we can define the cost function in the same way with $\phi$ replaced by $\hat{\phi}$. In the case where $\phi$ is unknown and allowed to vary over time, we define
\begin{align*}
    C(z_{s:t}) = \min_{\theta,\phi} - \sum_{i = s}^t \left\{ \frac{y_i\gamma_i - b(\gamma_i)}{w^{-1} \phi} + c(y_i, \phi) \right\},\quad g(\nabla b(\gamma_i)) = x_i^\top \theta.
\end{align*}

\subsection{Logistic regression}
\label{subsec:logistic regression model}

Consider the following logistic regression model:
\[
  y_t \sim \mathrm{Bernoulli} \left(\frac{1}{1+\exp(-x_t^\top \theta_t)}\right),
  \quad x_t\sim\mathcal{N}_p(0,I_p),\quad 1\le t\le T.
\]
Set $T=500$, $p=4$, and the true change point location at 300. The code below demonstrates how the \fct{fastcpd.binomial} function can be utilized to detect change points in this scenario.
\begin{Schunk}
\begin{Sinput}
R> result <- fastcpd.binomial(cbind(y, x))
R> summary(result)
\end{Sinput}
\begin{Soutput}
Call:
fastcpd.binomial(data = cbind(y, x))

Change points:
302

Cost values:
124.4554 54.26381

Parameters:
   segment 1 segment 2
1 -0.9260182 2.1294962
2 -1.6033835 2.7583247
3  1.0343338 2.3818010
4  0.3653870 0.7261152
\end{Soutput}
\end{Schunk}

\subsection{Poisson regression}
\label{subsec:poisson regression model}

Consider the following Poisson regression model:
\[
  y_t \sim \mathrm{Poisson}(\exp(x_t^\top \theta_t)),\quad
  x_t\sim\mathcal{N}_d(0,I_d),\quad 1\le t\le T.
\]
We set $T=1100$, $d=3$, and the true change point locations at
$500$, $800$ and $1000$ with the coefficients $\theta_t$ for each segment defined as
\begin{alignat*}{4}
  \theta_t &= \theta_0 &1\le t&\le 500, \\
  \theta_t &= \theta_0 + \delta & 501 \le t &\le 800, \\
  \theta_t &= \theta_0 & 801 \le t &\le 1000, \\
  \theta_t &= \theta_0 - \delta \quad & 1001 \le t &\le 1100,
\end{alignat*}
where $\delta_0=(1, 0.3, -1)$ and $\delta$ is a fixed random vector with
each element $delta_i$ ($i=1,2,3$) obtained from standard normal distribution.
The \fct{fastcpd.poisson} function can be used to detect change points in the coefficients of Poisson regression. Note that we set \code{epsilon} to be $10^{-5}$ to ensure that the Hessian matrix is invertible in the updates in SeGD.
\begin{Schunk}
\begin{Sinput}
R> result <- fastcpd.poisson(cbind(y, x), epsilon = 1e-5)
R> summary(result)
\end{Sinput}
\begin{Soutput}
Call:
fastcpd.poisson(data = cbind(y, x), epsilon = 1e-05)

Change points:
498 805 1003

Cost values:
230.0866 190.1381 82.77324 38.45199

Parameters:
  segment 1  segment 2  segment 3  segment 4
1  1.020002  0.6391880  1.0424108  1.4451928
2  0.275458 -0.2585056  0.2620085  0.9910079
3 -1.048875 -0.5768481 -0.9632918 -1.4354638
\end{Soutput}
\end{Schunk}

\section{Time series models}

Change point analysis is often concerned with data collected sequentially over time. It is, therefore, crucial to take into account the temporal dependence among the observations. In the subsequent sections, we shall discuss the usage of the \pkg{fastcpd} package for detecting change points in several stylized time series models, including AR, ARMA, GARCH, and VAR models.

\subsection{AR(\texorpdfstring{$p$}{p}) models}
\label{sec:autoregressive model}

We begin with the
autoregressive models of order $p$ (or the AR($p$) model) defined as
\begin{align*}
 x_t = \sum_{i = 1}^p \phi_i x_{t - i} + \epsilon_t,
\end{align*}
where $\phi_1, \ldots, \phi_p$ are the (autoregressive) coefficients and $\epsilon_t$ is a white noise with variance $\sigma^2$. As the AR($p$) model is an obvious extension of the linear regression models, we can borrow the techniques introduced in Section \ref{sec:changes-in-linear-models} for linear models. Specifically, we define $z_t=(x_t,x_{t-1},\dots,x_{t-p})$ and consider the cost function over the segment $z_{s:t}$ defined as
\begin{align*}
C(z_{s:t})=\min_{\theta}\left( \frac{1}{2}\log(2\pi\hat{\sigma}^2) +\sum_{i = s}^t\frac{(x_i -  x_{i-1:i-p}^\top \theta)^2}{2\hat{\sigma}^2}\right),
\end{align*}
where $x_{i-1:i-p}=(x_{i-1},\dots,x_{i-p})^\top$
and $\hat{\sigma}^2$ is the generalized Rice estimator developed in Section \ref{subsec: variance-estimation-with-change points}. The \fct{fastcpd.ar} function implements this method. To illustrate its usage, we consider an AR(3) model defined as
\begin{align*}
  x_t &= 0.6 x_{t - 1} - 0.2 x_{t - 2} + 0.1 x_{t - 3} + \epsilon_t
    \qquad  1\leq t \leq 600, \\
  x_t &= 0.3 x_{t - 1} + 0.4 x_{t - 2} + 0.2 x_{t - 3} + \epsilon_t
    \qquad  601\leq t\leq 1000,
\end{align*}
where $\epsilon_t \sim \mathcal{N}(0, 3^2)$. In this example, the \fct{fastcpd.ar} function detects a structural break at $t=614$. The \fct{summary} function returns the estimated parameters for each segment as well as the corresponding cost values.
\begin{Schunk}
\begin{Sinput}
R> result <- fastcpd.ar(ar3_data, order = 3)
R> summary(result)
\end{Sinput}
\begin{Soutput}
Call:
fastcpd.ar(data = ar3_data, order = 3)

Change points:
614

Cost values:
2743.759 2028.588

Parameters:
    segment 1 segment 2
1  0.57120256 0.2371809
2 -0.20985108 0.4031244
3  0.08221978 0.2290323
\end{Soutput}
\end{Schunk}
\begin{figure}
\centering
\begin{Schunk}
\begin{Sinput}
R> plot(result)
\end{Sinput}
\end{Schunk}
\includegraphics{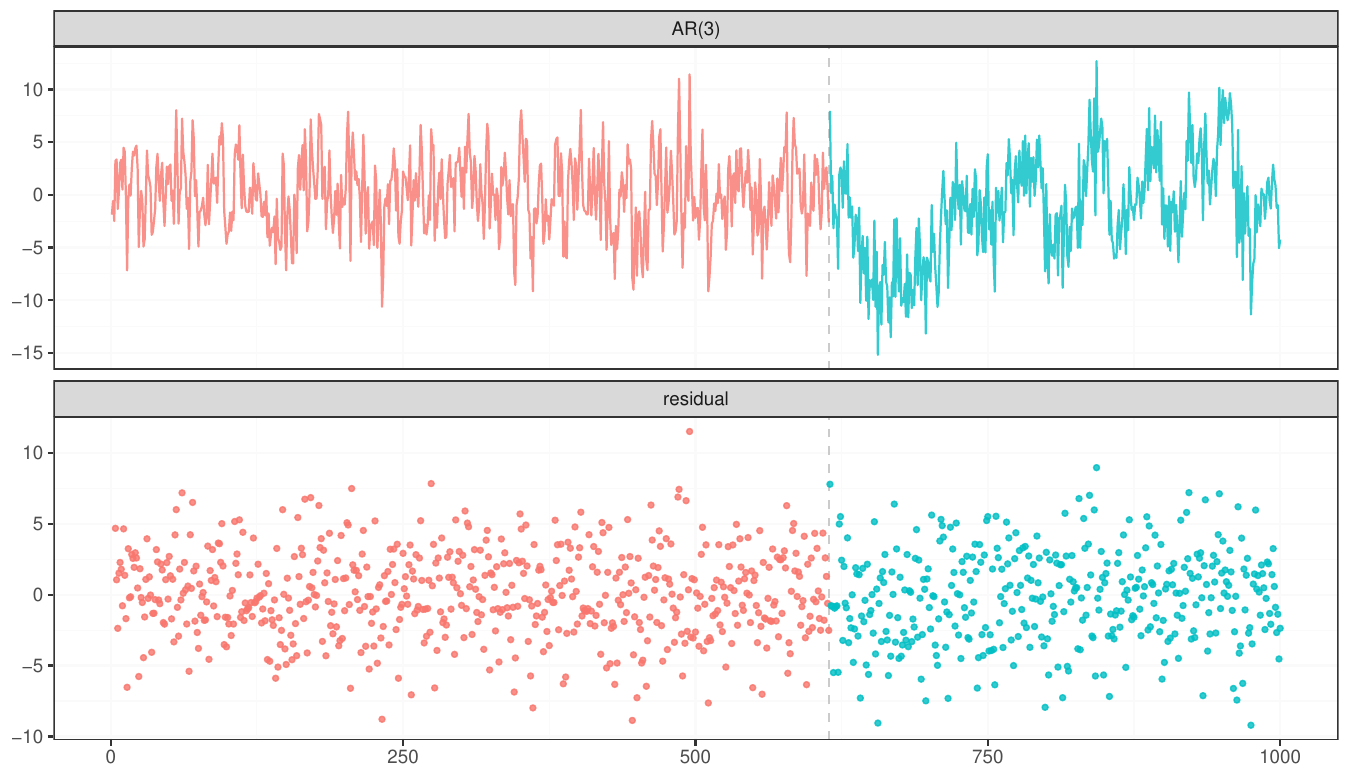}
\caption{Plots of the time series and the residuals from the fitted model together with the estimated change point location marked by the vertical grey line.
}
\label{fig:ar3-example-plot}
\end{figure}

\subsection{ARMA(\texorpdfstring{$p$}{p}, \texorpdfstring{$q$}{q}) models}
\label{sec:autoregressive moving average model}

The ARMA (Autoregressive Moving Average) model is a classical statistical model used in time series analysis, which generalizes and combines the AR model and the Moving Average (MA) model. A time series $\{x_t\}$ is an ARMA$(p, q)$ process if it is stationary and
\begin{align}
  x_t = \sum_{i = 1}^p \phi_i x_{t - i} + \sum_{i = 1}^q \psi_i \epsilon_{t - i}
        + \epsilon_t, \label{armamodeldefinition}
\end{align}
where $\boldsymbol{\phi} = (\phi_1, \phi_2, \ldots, \phi_p)^\top$ and $\boldsymbol{\psi} = (\psi_1, \psi_2, \ldots, \psi_q)^\top$ are the AR and MA coefficients respectively and $\epsilon_t$ is a white noise sequence with variance $\sigma^2$.

Here we consider a time-varying ARMA model where $\boldsymbol{\phi}$, $\boldsymbol{\psi}$ and $\sigma$ can change over time, i.e.,
\begin{align*}
    x_t = \sum_{i = 1}^p \phi_{i, t} x_{t - i} + \sum_{i = 1}^q \psi_{i, t} \epsilon_{t - i} + \epsilon_t,
\end{align*}
where $\boldsymbol{\phi}_t = (\phi_{1, t},\ldots,\phi_{p,t})$, $\boldsymbol{\psi}_t = (\psi_{1,t},\ldots,\psi_{q,t})$ and $\epsilon_t$ is a white noise sequence with variance $\sigma_t^2$.

\textbf{Coefficient changes.}
Suppose we have a data sequence generated from the ARMA model with time-varying AR and MA coefficients and $\sigma_t^2 = \sigma^2$ fixed but unknown. Assume the ARMA model is invertible, and we can represent the ARMA model with an AR($\infty$) model. \citep{shumway2000time}.
Here, we adapt the Rice estimator in Section~\ref{subsec: variance-estimation-with-change points} to estimate $\sigma^2$:
\begin{enumerate}
    \item We approximate the time-varying ARMA($p$, $q$) model by a time-varying AR($p'$) model with $p' \gg p$, where the order $p'$ is chosen using AIC or BIC described below.
    \item We estimate $\sigma^2$ by the generalized Rice estimator developed in Section \ref{subsec: variance-estimation-with-change points} with the linear model defined as
    \begin{align*}
        y_t &= \boldsymbol{x}_t^\top \theta_t + \epsilon_t,\quad \epsilon \sim \mathcal{N}(0, \sigma^2),\quad t = 1,\ldots,T,
    \end{align*}
    where $y_t = x_t$, $\boldsymbol{x}_t = (x_{t - 1}, \ldots, x_{t - p'})$ and $\theta_t = (\phi_{1, t}, \ldots, \phi_{p', t})$ is assumed to be piece-wise constant.
\end{enumerate}
The order $p'$ for the AR model (used to approximate the ARMA model) can be obtained by minimizing the AIC or BIC \citep{hannan1984method} defined as:
\begin{align*}
    \mathrm{AIC}(p') = \log(\hat{\sigma}^2) + 2 p' / T, \qquad \mathrm{BIC}(p') = \log(\hat{\sigma}^2) + p' \log(T) / T.
\end{align*}
We define the cost function for the data segment $x_{s:t}$ utilizing $\hat{\sigma}^2$ obtained through the above procedure as follows:
\begin{align*}
    C(x_{s:t}) = \min_{\boldsymbol{\phi}, \boldsymbol{\psi}} \frac{t - s + 1}{2}
  \left \lbrace
    \log(2 \pi) + \log(\hat{\sigma}^2) \right \rbrace +
    \frac{1}{2 \hat{\sigma}^2} \sum_{i = s}^t (x_i-\boldsymbol{\phi}^\top x_{i-1:i-p} - \boldsymbol{\psi}^\top \epsilon_{i-1:i-q})^2.
\end{align*}

\textbf{Coefficient and variance changes.}
The \fct{fastcpd.arma} function implements the vanilla PELT when \code{vanilla_percentage}\footnote{See Section~\ref{subsec:advanced usage vanilla pelt} for more discussions about this parameter.} is set to be 1, where the cost value is defined as the negative log-likelihood which can be computed using \fct{forecast::Arima} \citep{forecastrpackage, forecastrpaper}.
To speed up the calculation, the \fct{fastcpd.arma} function also implements the SeGD when \code{vanilla_percentage < 1} with the following cost function. Specifically, assuming $\epsilon_{0:1-q} = \mathbf{0}$ and $x_{0:1-p}=\mathbf{0}$, we define the cost value for the data segment $x_{s:t}$ as
\begin{align*}
  C(x_{s:t}) = \min_{\boldsymbol{\phi}, \boldsymbol{\psi}, \sigma^2} \frac{t - s + 1}{2}
  \left \lbrace
    \log(2 \pi) + \log(\sigma^2) \right \rbrace +
    \frac{1}{2 \sigma^2} \sum_{i = s}^t (x_i-\boldsymbol{\phi}^\top x_{i-1:i-p} - \boldsymbol{\psi}^\top \epsilon_{i-1:i-q})^2,
\end{align*}
where $\epsilon_t = x_t - \boldsymbol{\phi}^\top x_{t-1:t-p} - \boldsymbol{\psi}^\top \epsilon_{t-1:t-q}$ can be obtained recursively.
The cost function defined above is closely related to the conditional likelihood and is preferred over the exact likelihood due to its simpler form, which improves computational efficiency. The explicit forms of the corresponding gradient and Hessian can be found in Appendix~\ref{sec:gradient and hessian of qmle for arma(p, q) model}.

\begin{figure}
\centering
\begin{Schunk}
\begin{Sinput}
R> plot(result)
\end{Sinput}
\end{Schunk}
\includegraphics{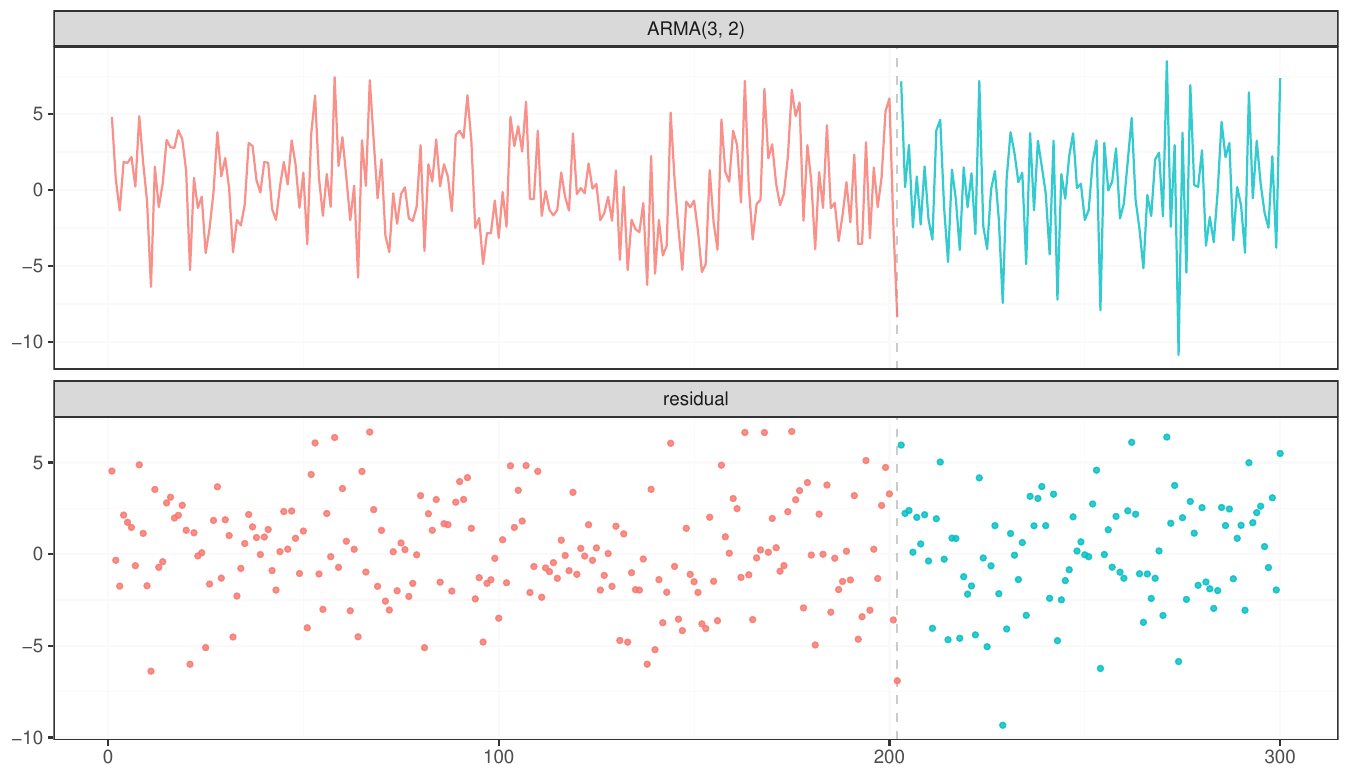}
\caption{Plots of the time series and residuals from the fitted ARMA model with the estimated change point location, which is marked with a vertical grey line.}
\label{fig:arma-32-model-plot}
\end{figure}

We now illustrate the use of the \fct{fastcpd.arma} function using SeGD with the above cost function. Consider the following data-generating process
\begin{align*}
  x_t &= 0.1 x_{t - 1} - 0.3 x_{t - 2} + 0.1 x_{t - 3} + 0.1 \epsilon_{t - 1} +
    0.5 \epsilon_{t - 2} + \epsilon_t \qquad 1 \leq t \le 200,\\
  x_t &= 0.3 x_{t - 1} + 0.1 x_{t - 2} - 0.3 x_{t - 3} - 0.6 \epsilon_{t - 1} -
    0.1 \epsilon_{t - 2} + \epsilon_t \qquad 201\leq t\leq 300,
\end{align*}
where $\epsilon_t \sim \mathcal{N}(0, 1)$ for $t = 1, 2, \ldots, 300$.
We project the updated parameters in each step onto the following domains when performing SeGD for better convergence (see \eqref{projectedupdatestep} for the projection step in SeGD) by specifying the \code{lower} and \code{upper} parameters in the function:
\begin{align*}
  \phi_i &\in (-1, 1), \qquad i = 1, 2, \ldots, p, \\
  \psi_j &\in (-1, 1), \qquad j = 1, 2, \ldots, q, \\
  \sigma^2 &> 0.
\end{align*}
An additional line search is performed at each updating step to find the optimal step size by specifying the \code{line_search} parameter (see Section~\ref{subsec:advanced usage line search} for detailed instructions on how to use line search).
\begin{Schunk}
\begin{Sinput}
R> result <- fastcpd.arma(
+    data = arma32_data,
+    order = c(3, 2),
+    segment_count = 3,
+    lower = c(-1, -1, -1, -1, -1, 1e-10),
+    upper = c(1, 1, 1, 1, 1, Inf),
+    line_search = c(1, 0.1, 1e-2)
+  )
R> summary(result)
\end{Sinput}
\begin{Soutput}
Call:
fastcpd.arma(data = arma32_data, order = c(3, 2), segment_count = 3,
    lower = c(-1, -1, -1, -1, -1, 1e-10), upper = c(1, 1, 1,
        1, 1, Inf), line_search = c(1, 0.1, 0.01))

Change points:
202

Cost values:
498.3128 241.8409

Parameters:
     segment 1  segment 2
1  0.001958689  0.7431369
2 -0.819059235 -0.0443085
3  0.187015202 -0.3922486
4  0.207797827 -1.2583378
5  0.999989911  0.4421605
6  8.054595972  8.6103689
\end{Soutput}
\end{Schunk}
The residuals from the fitted model with the estimated change point are approximately white noise, as shown in Figure~\ref{fig:arma-32-model-plot}.

\subsection{GARCH(\texorpdfstring{$p$}{p}, \texorpdfstring{$q$}{q}) models}
\label{sec:garch model}

Generalized Autoregressive Conditional
Heteroskedasticity (GARCH) models play a pivotal role in modeling volatility and capturing the time-varying nature of variances within a time series. While ARIMA models effectively capture the autocorrelation and trend components of time series data, GARCH models extend their capabilities by addressing volatility clustering, a common characteristic in financial markets.

A (time-varying) GARCH($p$, $q$) model can be expressed in the following form:
\begin{align*}
&x_t =\sigma_t \epsilon_t,\\
&\sigma_t^2 = \omega + \sum_{i = 1}^p \alpha_{i,t} x_{t - i}^2 +
    \sum_{j = 1}^q \beta_{j,t} \sigma_{t - j}^2,
\end{align*}
for $t = 1,\ldots, T$, where $\sigma_t^2$ is the conditional variance of the time series at time $t$, $\omega>0$ is the constant term, $\alpha_{i,t}\geq 0$ and $\beta_{j,t}\geq 0$ are the autoregressive and moving average parameters respectively
which are assumed to be piece-wise constant over time, and $\epsilon_t$ represents the white noise term at time $t$.

Consider a sequence of observations generated from a
GARCH(1, 1) model with a change point located at $t = 200$:
\begin{alignat*}{4}
    x_t &= \sigma_t \epsilon_t &\qquad &\epsilon_t \sim \mathcal{N}(0, 1), \\
  \sigma_t^2 &=
    20 + 0.5 x_{t - 1}^2 + 0.1 \sigma_{t - 1}^2 &\qquad &1\leq t \le 200, \\
  \sigma_t^2 &= 1 + 0.1 x_{t - 1}^2 + 0.5 \sigma_{t - 1}^2 &\qquad  & 201\leq t \leq  400.
\end{alignat*}
The \fct{fastcpd.garch} function is specifically designed for detecting change points in GARCH models with the vanilla PELT algorithm utilizing \fct{tseries::garch} to obtain the negative log-likelihood as the cost value. In this example, the \fct{fastcpd.garch} function detects a change point at $t = 205$, which is fairly close to the true change point at $t=200.$
\begin{Schunk}
\begin{Sinput}
R> result <- fastcpd.garch(garch11_data, c(1, 1))
R> summary(result)
\end{Sinput}
\begin{Soutput}
Call:
fastcpd.garch(data = garch11_data, order = c(1, 1))

Change points:
205

Cost values:
450.6641 179.6463
\end{Soutput}
\end{Schunk}
\begin{figure}
\centering
\begin{Schunk}
\begin{Sinput}
R> plot(result)
\end{Sinput}
\end{Schunk}
\includegraphics{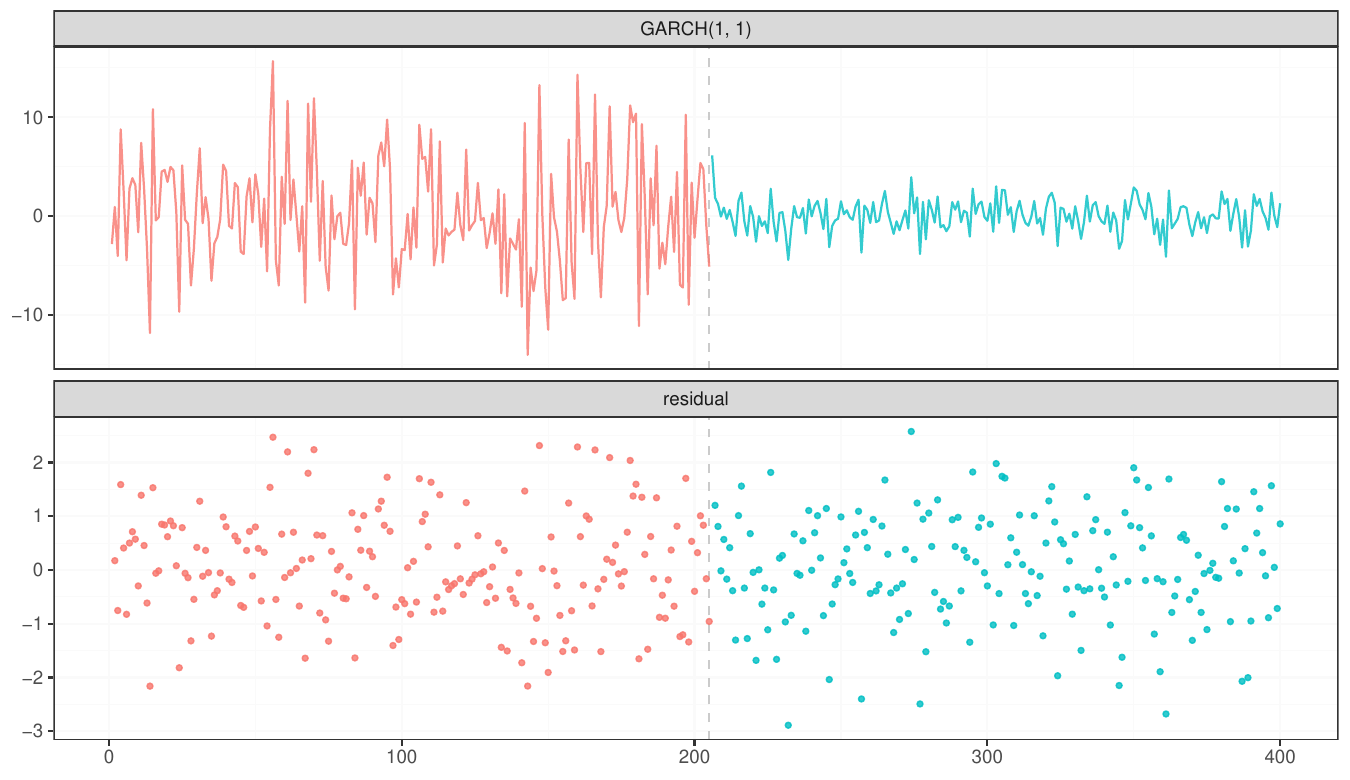}
\caption{Plots of the time series and residuals from the fitted GARCH model with the estimated change point location marked with a vertical grey line.}
\label{fig:garch-model-plot}
\end{figure}

\subsection{VAR(\texorpdfstring{$p$}{p}) models}
\label{sec:vector autoregressive model}

Vector Autoregressive (VAR) models extend the concept of autoregressive models to multivariate time series. In this section, we demonstrate how the fastcpd package can be utilized to detect structural breaks in the VAR($p$) models. A VAR($p$) model for a multivariate time series $\{\boldsymbol{x}_t\}$ of dimension $d$ is defined as follows:
\[
  \boldsymbol{x}_t = \sum_{i = 1}^p A_i \boldsymbol{x}_{t - i} +
    \boldsymbol{\epsilon}_t,
\]
where $\boldsymbol{x}_t\in\mathbb{R}^d$ is a column vector representing the observations at time $t$, $A_i\in\mathbb{R}^{d\times d}$ are the coefficient matrices, and $\boldsymbol{\epsilon}_t$ is a vector of white noise.
Here we consider a time-varying VAR model where $A_i$ can change over time, i.e.,
\begin{align*}
    \boldsymbol{x}_t = \sum_{i = 1}^p A_{i, t} \boldsymbol{x}_{t - i} + \boldsymbol{\epsilon}_t,\quad \boldsymbol{\epsilon}_t \sim \mathcal{N}(0, \Sigma),\quad t = 1\ldots,T,
\end{align*}
where $\Sigma$ is fixed but unknown. We assume that the coefficient matrices $\{(A_{1,t}, \ldots, A_{p,t})\}_{t = 1}^T$ are piece-wise constant over time.
We define $z_t=(\boldsymbol{x}_t,\boldsymbol{x}_{t-1},\dots,\boldsymbol{x}_{t-p})$ and consider the cost function over the segment $z_{s:t}$ as
\begin{align*}
C(z_{s:t})=&\min_{A_1,\dots,A_p}\frac{1}{2}\sum_{i = s}^t \left(\boldsymbol{x}_i-\sum_{i = 1}^p A_{i} \boldsymbol{x}_{t - i}\right) ^\top\hat{\Sigma}^{-1}\left(\boldsymbol{x}_i-\sum_{i = 1}^p A_{i} \boldsymbol{x}_{t - i}\right)
\\&+ \frac{t-s+1}{2}\{d\log(2\pi) + \log(\lvert \hat{\Sigma} \rvert)\},
\end{align*}
where $\lvert\ \cdot\ \rvert$ denotes the matrix determinant and $\hat{\Sigma}$ is an extension of the generalized Rice estimator developed in Section \ref{subsec: variance-estimation-with-change points} to multi-response linear models. The \fct{fastcpd.var} function implements this method. To illustrate its usage, we consider a VAR(2) model defined as
\begin{align*}
  \boldsymbol{x}_t &= A_1 \boldsymbol{x}_{t - 1} + A_2 \boldsymbol{x}_{t - 2} +
    \boldsymbol{\epsilon}_t \qquad  1\leq t \le 200, \\
  \boldsymbol{x}_t &= B_1 \boldsymbol{x}_{t - 1} + B_2 \boldsymbol{x}_{t - 2} +
    \boldsymbol{\epsilon}_t \qquad 201 \leq t \leq  300,
\end{align*}
where $\boldsymbol{\epsilon}_t \sim \mathcal{N}(\boldsymbol{0}, I_2)$ and
$A_1$, $A_2$, $B_1$, and $B_2$ are the coefficient matrices defined as
\begin{align*}
    A_1 = \begin{pmatrix}
        -0.3 & -0.5 \\
        0.6 & 0.4
    \end{pmatrix}, \quad A_2 = \begin{pmatrix}
        0.2 & 0.2 \\
        0.2 & -0.2
    \end{pmatrix}, \quad B_1 = \begin{pmatrix}
        0.3 & 0.1 \\
        -0.4 & -0.5
    \end{pmatrix}, \quad B_2 = \begin{pmatrix}
        -0.5 & -0.5 \\
        -0.2 & 0.2
    \end{pmatrix}.
\end{align*}
\begin{Schunk}
\begin{Sinput}
R> result <- fastcpd.var(var_data, 2)
R> summary(result)
\end{Sinput}
\begin{Soutput}
Call:
fastcpd.var(data = var_data, order = 2)

Change points:
202

Cost values:
555.4247 283.8178
\end{Soutput}
\end{Schunk}

\section{Custom models}
\label{sec:custom cost function}

This section will explore how to use custom cost functions in the \pkg{fastcpd} package. The \fct{fastcpd} function allows users to provide their own cost functions/values together with the gradient and Hessian information via the \code{cost}, \code{cost_gradient}, and \code{cost_hessian} parameters. This feature
enables users to tailor the change model to specific objectives/needs and hence widens the scope of the package. Below, we consider an example from robust statistics to illustrate the usage of this feature.

Huber regression is a type of robust regression method, which is less sensitive to outliers than the
ordinary least squares \citep{huber1992robust}. Suppose we have
a data sequence $\{z_t = (x_t,y_t)\}_{t = 1}^T$, where $x_t \in \mathbb{R}^d$ is the
regressor and $y_t \in \mathbb{R}$ is the response.
The cost value in the Huber regression model for the segment $z_{s:t}$ is defined as
\begin{align}
  C(z_{s:t}) = \min_{\theta} \sum_{i = s}^t \rho_{\delta}(y_i - x_i^\top \theta), \label{eq:huberloss}
\end{align}
with the Huber loss $\rho_{\delta}(u)$ given by
\begin{align}
  \rho_{\delta}(u) = \begin{cases}
    \frac{1}{2} u^2 & \text{if } |u| \le \delta, \\
    \delta |u| - \frac{1}{2} \delta^2 & \text{otherwise},
  \end{cases}
\end{align}
for some $\delta>0$.
Here, we follow similar settings as in \pkg{CVXR} \citep{cvxr2020} where the covariates are generated from a multivariate normal distribution, and the response has a $5\%$ chance of being flipped. The true regression coefficients for each segment are randomly generated from multivariate normal distributions with different means. More precisely, suppose we observe a sequence of data points generated from the following model
\begin{align*}
  \tilde{y}_t &= \boldsymbol{x}_t^\top \boldsymbol{\theta}_1 + \epsilon_t
    \qquad 1\leq t \le 400, \\
  \tilde{y}_t &= \boldsymbol{x}_t^\top \boldsymbol{\theta}_2 + \epsilon_t
    \qquad 401\leq t \leq 700, \\
  \tilde{y}_t &= \boldsymbol{x}_t^\top \boldsymbol{\theta}_3 + \epsilon_t
    \qquad 701\leq t\leq 1000, \\
  y_t &= \begin{cases}
    \tilde{y}_t & \text{with probability } 0.95, \\
    -\tilde{y}_t & \text{with probability } 0.05,
  \end{cases}
\end{align*}
where $\boldsymbol{x}_t \sim \mathcal{N}(\boldsymbol{0}, I_d)$,
$\epsilon_t \sim \mathcal{N}(0, 1)$ and $d = 5$. Here
$\boldsymbol{\theta}_1$, $\boldsymbol{\theta}_2$ and $\boldsymbol{\theta}_3$
are three $p$-dimensional random vectors generated from the multivariate normal
distributions with the means vectors $(0, 0, 0, 0, 0)$, $(5, 5, 0, 0, 0)$,
$(9, 9, 0, 0, 0)$ and the covariance matrix $I_p$ respectively. Define the
cost function in Eq~\eqref{eq:huberloss} as \code{huber_loss}, with the
corresponding gradient and Hessian functions as \code{huber_loss_gradient} and
\code{huber_loss_hessian} respectively. Then we can use the \fct{fastcpd}
function to detect the change points in the Huber regression model.
\begin{Schunk}
\begin{Sinput}
R> fastcpd(
+    formula = y ~ . - 1,
+    data = huber_data,
+    cost = huber_loss,
+    cost_gradient = huber_loss_gradient,
+    cost_hessian = huber_loss_hessian
+  )@cp_set
\end{Sinput}
\begin{Soutput}
[1] 418 726
\end{Soutput}
\end{Schunk}

\section{Real data analysis} \label{sec: real data analysis}

We present four real data sets to demonstrate the usefulness of the package under various change point models. These include: (i) the well-log data set from \cite{Ruanaidh1996}, which has been widely utilized to test various mean change point detection algorithms \citep{fearnhead2006exact, haynes2014efficient, fearnhead2019changepoint, van2020evaluation}; (ii) the micro-array aCGH data from \cite{Stransky2006}, included in the \pkg{ecp} package \citep{james2013ecp}, and utilized by \cite{bleakley2011group} to demonstrate the group fused lasso algorithm; (iii) the daily prices of Bitcoin in USD\footnote{Obtained from \url{https://www.blockchain.com/explorer/charts/market-price}}; and (iv) an \proglang{R} built-in data set regarding Road Casualties in Great Britain from 1969 to 1984.

\subsection{Well-log data} \label{subsec:well-log data}
\begin{figure}
\centering
\begin{Schunk}
\begin{Sinput}
R> result <- fastcpd.mean(well_log, trim = 0.002)
R> plot(result)
\end{Sinput}
\end{Schunk}
\includegraphics{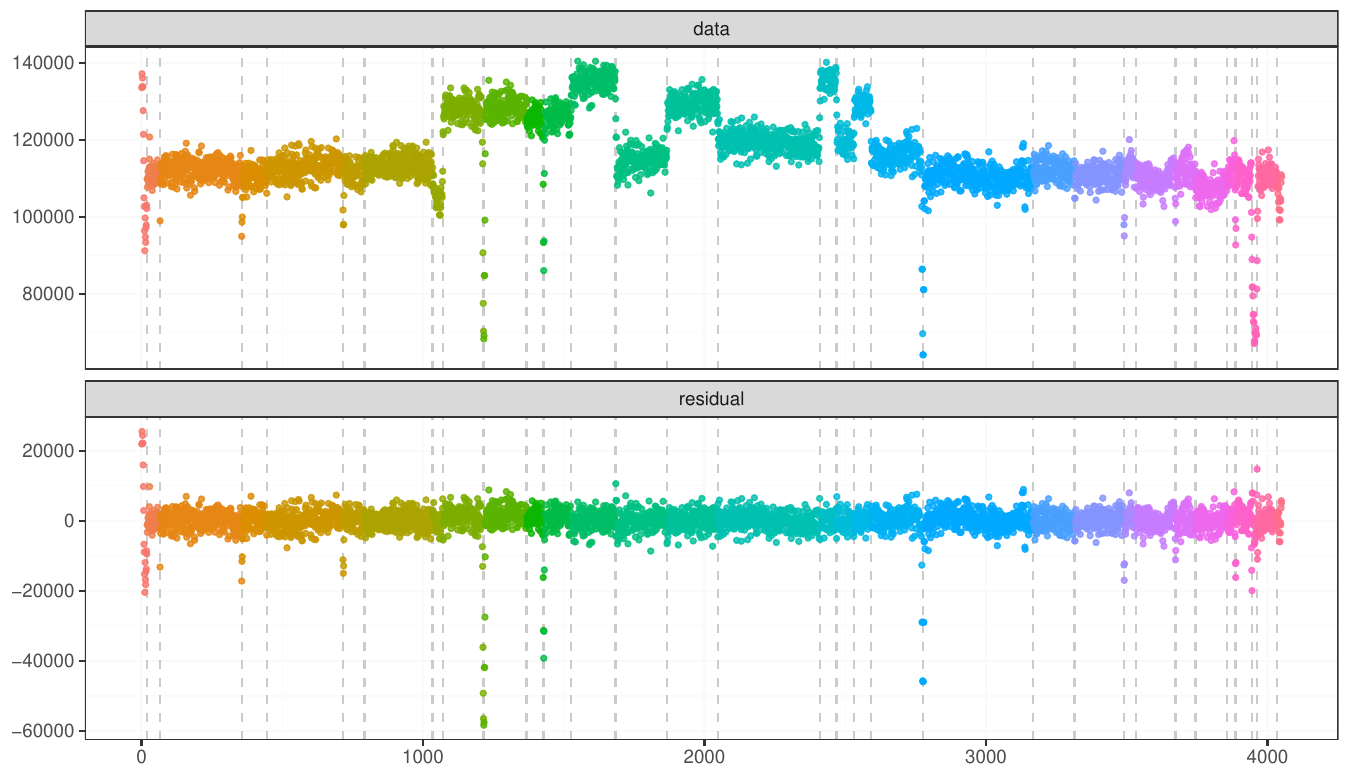}
\caption{The Well-log data with the detected change points marked by the vertical grey lines.}
\label{fig:well-log-data-summary}
\end{figure}

This dataset consists of well-log data comprising 4050 measurements capturing the nuclear-magnetic response of subsurface rocks. The measurements were obtained using a probe lowered into a borehole, providing a discrete-time profile of the geological composition. The dataset's inherent signal displays a piecewise constant structure, where each constant segment corresponds to a distinct rock type characterized by uniform physical properties. This dataset is particularly relevant in the context of oil drilling, where identifying these change points is crucial for understanding shifts in geological composition. Further insights can be found in the work of \cite{fearnhead2003line}.
Figure~\ref{fig:well-log-data-summary} presents the change points detected using the \fct{fastcpd.mean} function.

\begin{figure}
\centering
\includegraphics{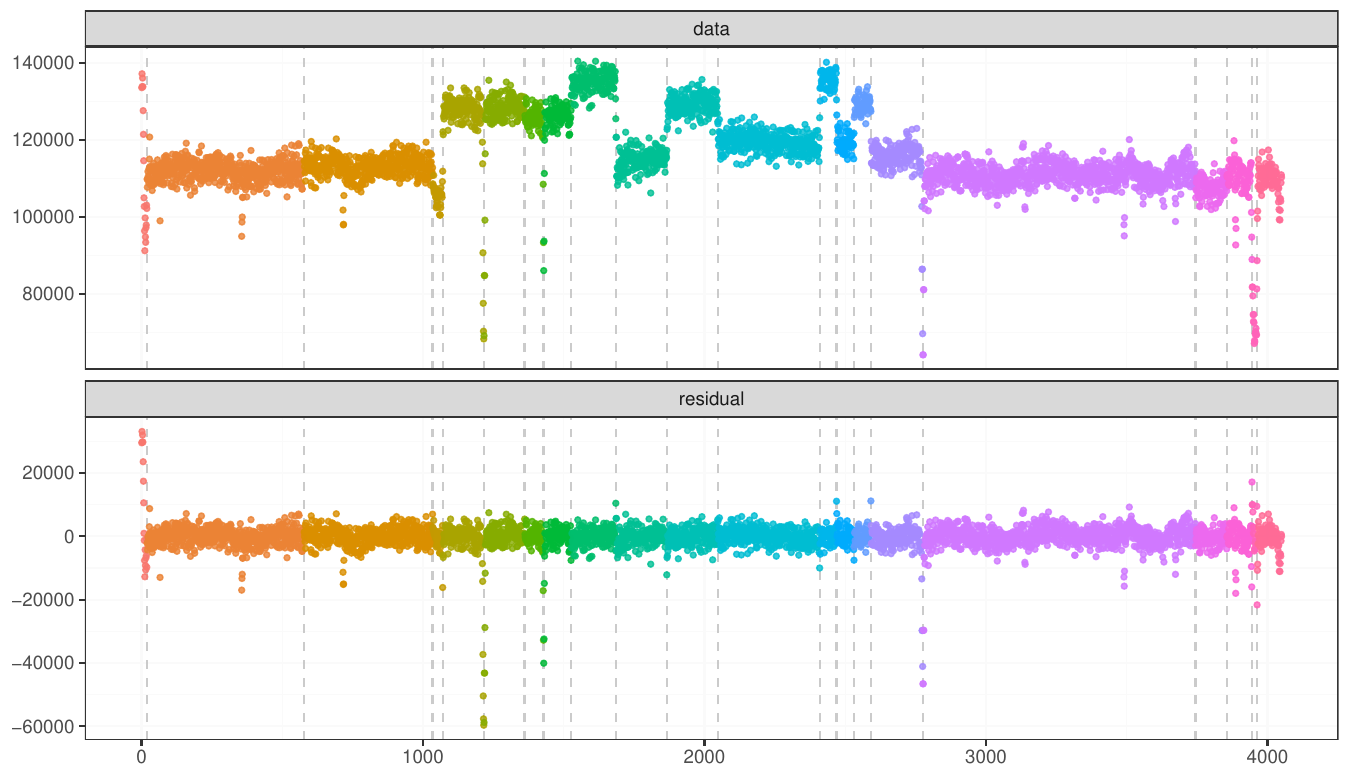}
\caption{The change points in the median detected by the \pkg{fastcpd} package for the
well-log data set.}
\label{fig:well-log-data-summary-larger-beta}
\end{figure}

Due to the potential outliers in the data set, it is reasonable to consider an algorithm or a cost function that is robust to outliers \citep{fearnhead2019changepoint}. To this end, we propose to detect the structural break in the median of the data.
As a motivation, suppose we have observed the data sequence $\{x_t\}_{t = 1}^T$ with $x_t = \mu_t + \epsilon_t$, where $\epsilon_t$ follows the Laplace distribution:
\[f(\epsilon;b) = \frac{1}{2b}\exp(-\epsilon/b),\quad t = 1,\ldots,T,\]
with the variance $2b^2$. We adopt the loss function from quantile regression and define the cost function for the data segment $x_{s:t}$ as
\[C(x_{s:t}) = \min_{\theta} \frac{1}{\hat{\sigma}} \sum_{i=s}^t \rho_{0.5}(x_i - \theta) = \min_{\theta} \frac{1}{2\sqrt{\hat{\sigma}^2}} \sum_{i=s}^t \lvert x_i - \theta \rvert = \frac{1}{2\sqrt{\hat{\sigma}^2}}\sum_{i=s}^t\lvert x_i-\tilde{x}_{s:t}\rvert\]
with $\hat{\sigma}^2$ serving as an estimate of the variance of $\epsilon_t$ and $\tilde{x}_{s:t}$ being the median of $x_{s:t}$. Here, $\hat{\sigma}^2$ is an extension of the Rice estimator for Laplace distributions.
Since $E\lvert \epsilon_{t+1} - \epsilon_t \rvert = 3b/2$ for $t = 1,\ldots,T-1$, the Rice estimator can be defined as
\begin{align*}
    \hat{\sigma}^2 = 2 \hat{b}^2 = 2 \left(\frac{2}{3(T-1)}\sum_{t = 1}^{T-1}\lvert x_{t+1}-x_t\rvert\right)^2.
\end{align*}
\begin{Schunk}
\begin{Sinput}
R> (sigma2 <- variance.median(well_log))
\end{Sinput}
\begin{Soutput}
[1] 5803645
\end{Soutput}
\begin{Sinput}
R> median_loss <- function(data) {
+    sum(abs(data - matrixStats::colMedians(data))) / sqrt(sigma2) / 2
+  }
R> result <- fastcpd(
+    formula = ~ x - 1,
+    data = cbind.data.frame(x = well_log),
+    cost = median_loss,
+    trim = 0.002
+  )
\end{Sinput}
\end{Schunk}
Figure~\ref{fig:well-log-data-summary-larger-beta} shows the result for testing the median change for the well-log data set. We can observe that the number of the detected change points becomes smaller, and it is worth noting that the change points detected in the mean change model due to the outliers at $t=356$, $t=717$, and $t=3490$ are no longer considered as change points in the median change model.

\begin{figure}
  \centering
  \includegraphics[scale=1]{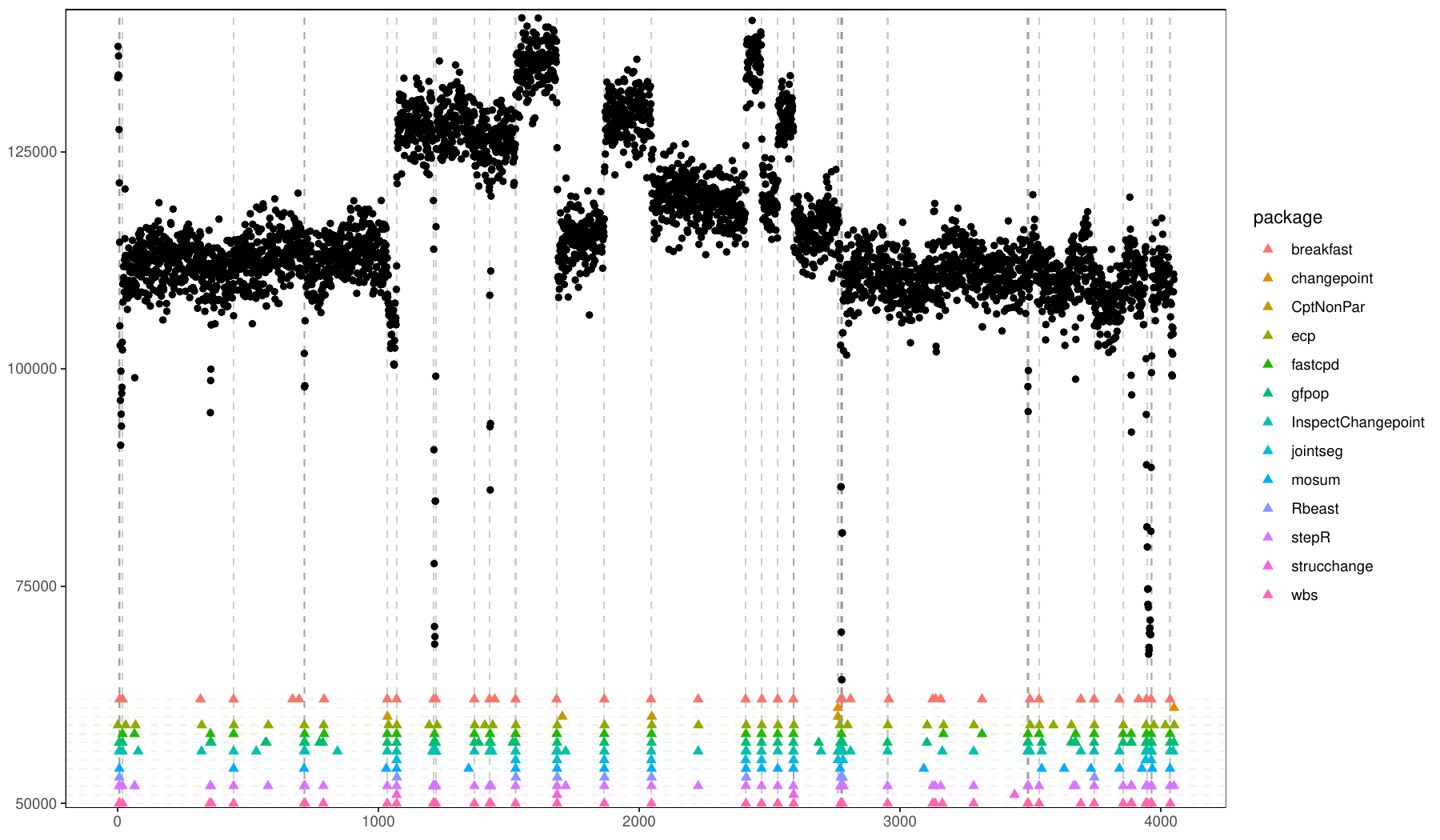}
  \caption{The figure presents a comprehensive comparison of change point detection results obtained from multiple packages on the well-log dataset, with the outcome of each package represented by distinct colors. Grey dashed vertical lines delineate the consensus change points identified by at least four packages, serving as a reference for the most commonly detected change points across the methods. The analysis executed by \fct{fastcpd.mean} was completed within a second. The exclusion of certain packages from this comparison is attributed to a range of reasons, including their absence from CRAN, incompatibility with the dataset in question, or the production of irrelevant change point locations.
  }
  \label{fig:efficiency comparison}
\end{figure}

\subsection{Micro-array aCGH data set} \label{subsec:micro-array acgh data set}

\begin{figure}
\centering
\begin{Schunk}
\begin{Sinput}
R> result <- fastcpd.mean(transcriptome$"10", trim = 0.005)
R> plot(result)
\end{Sinput}
\end{Schunk}
\includegraphics{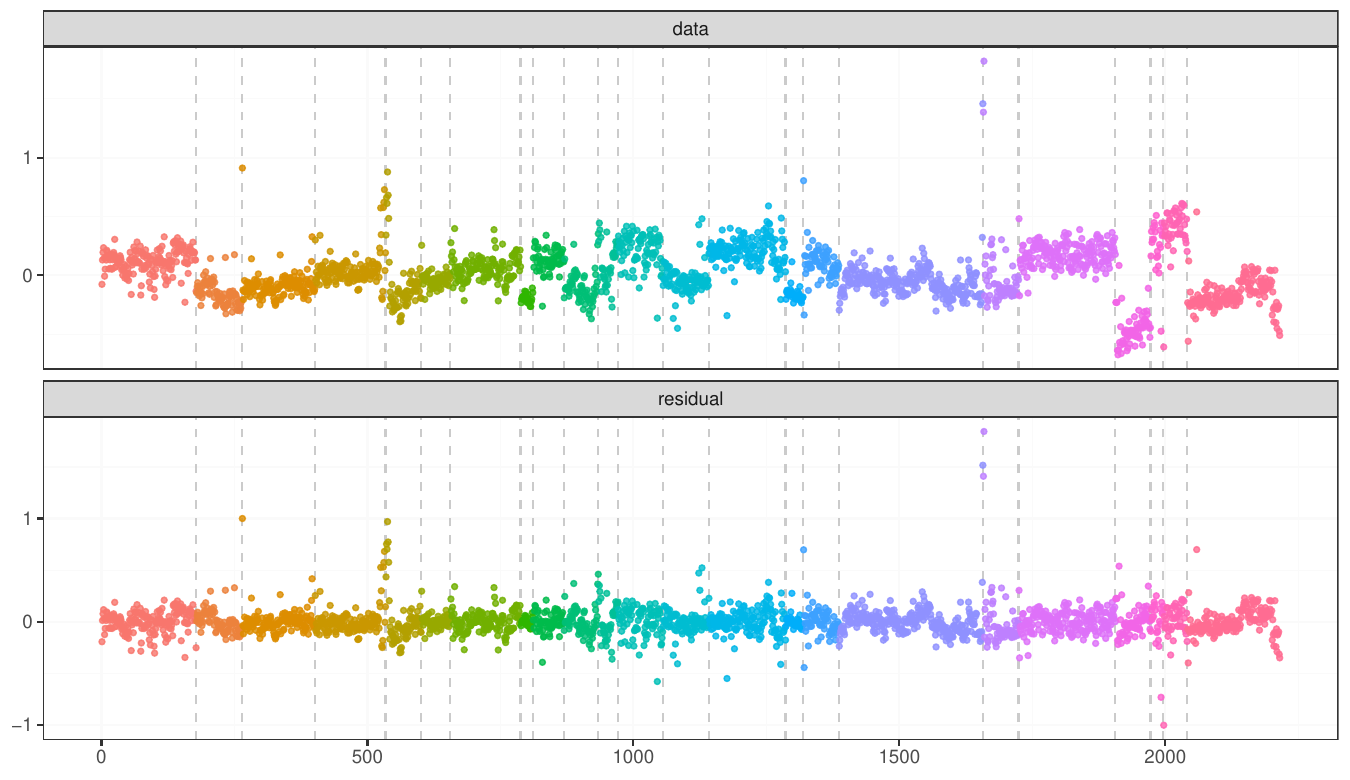}
\caption{Change points detected for individual 10 in the micro-array aCGH data
set.}
\label{fig:micro-array-acgh-data-summary}
\end{figure}

We analyze the aCGH dataset \citep{james2013ecp}, which consists of micro-array data from $43$ individuals with bladder tumors. As they have a shared medical condition, it is expected that the locations of change points would be almost identical across each microarray set. In this context, a change point represents a shift in copy-number that is assumed to remain constant within each segment \citep{james2013ecp}. The Group Fused Lasso (GFL) method, proposed by \cite{bleakley2011group}, is well-suited for detecting changes in mean. Figure~\ref{fig:micro-array-acgh-data-summary} illustrates the change points detected by the \pkg{fastcpd} package for individual 10.

\begin{figure}
\centering
\includegraphics{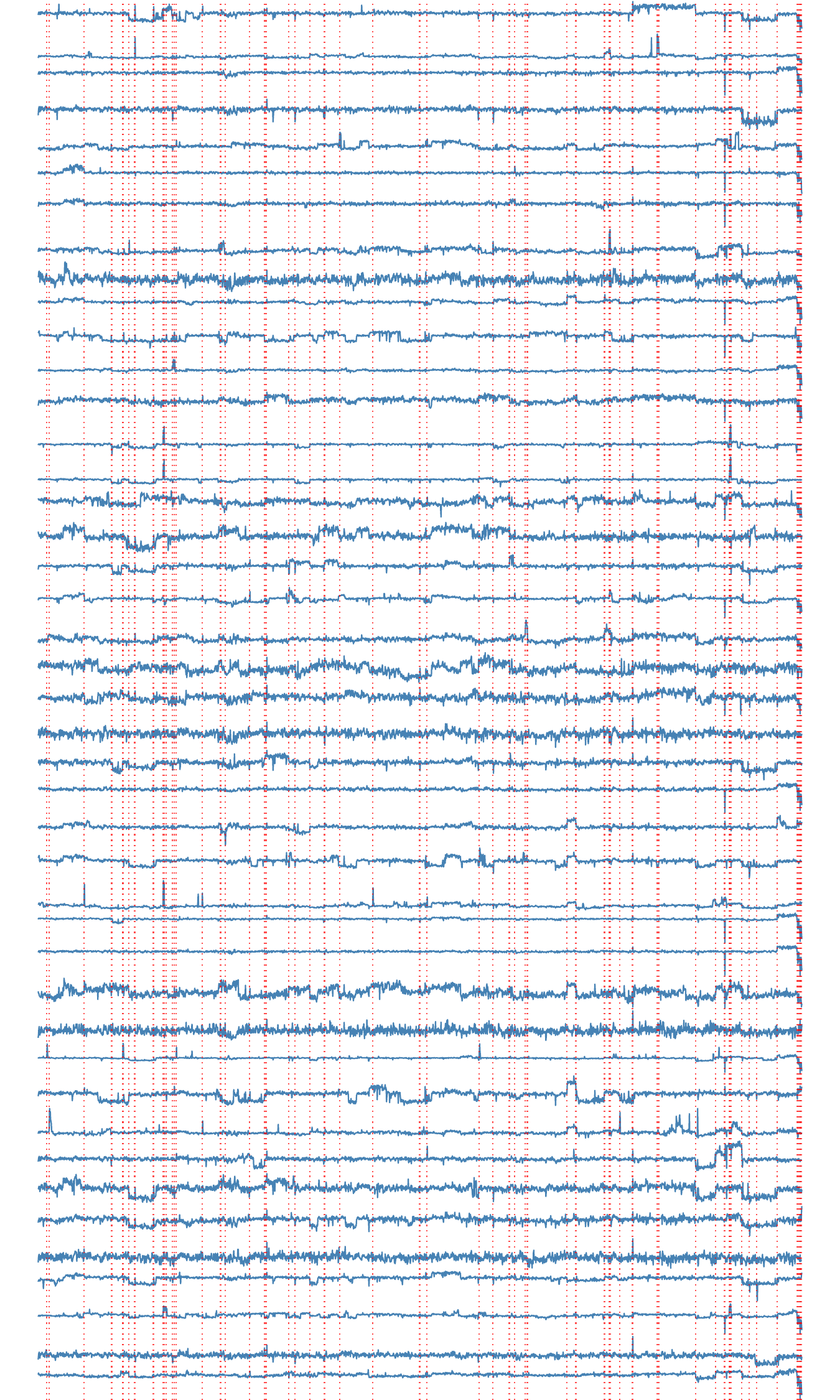}
\caption{Change points detected for all individuals in the micro-array aCGH data
  set.}
\label{fig:micro-array-acgh-data-summary-whole-data-set-plot}
\end{figure}

We further apply the \fct{fastcpd.mean} function to the entire data set, i.e., we aim to test the mean change of a multivariate sequence of dimension 43. As seen from Figure~\ref{fig:micro-array-acgh-data-summary-whole-data-set-plot}, most of the detected change points are shared by multiple patients.
\begin{Schunk}
\begin{Sinput}
R> result_all <- fastcpd.mean(transcriptome, trim = 0.0005)
\end{Sinput}
\end{Schunk}

\subsection{Bitcoin market price}
\label{subsec:bitcoin market price (USD)}

The Bitcoin market price is a representation of the average USD value of Bitcoin across major crypto exchanges. This dataset covers 1354 days, from January 2, 2009, to October 28, 2023, and shows the dynamic nature of Bitcoin's price. Here, we consider a GARCH(1,1) model \citep{katsiampa2017volatility} with potential change points in the parameters to understand the patterns of volatility of the Bitcoin return series.

\begin{figure}
\centering
\begin{Schunk}
\begin{Sinput}
R> result <- fastcpd.garch(
+    diff(log(bitcoin$price[600:900])), c(1, 1),
+    beta = "BIC", cost_adjustment = "BIC"
+  )
R> plot(result)
\end{Sinput}
\end{Schunk}
\includegraphics{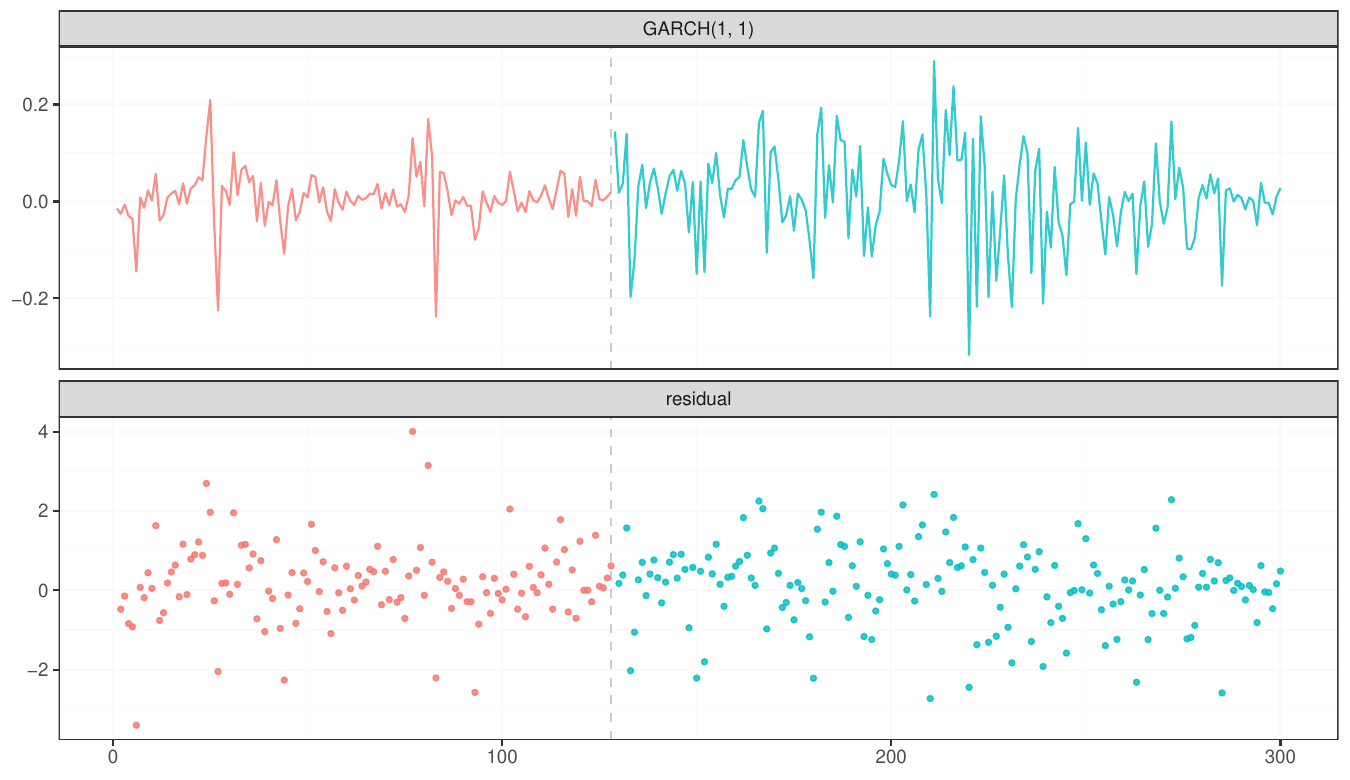}
\caption{Plots of Bitcoin return series and the residuals from the fitted GARCH model with a change point detected on
December 23, 2016, marked by the grey vertical line.}
\label{fig:bitcoin-market-price-summary}
\end{figure}

\begin{figure}
\centering
\includegraphics{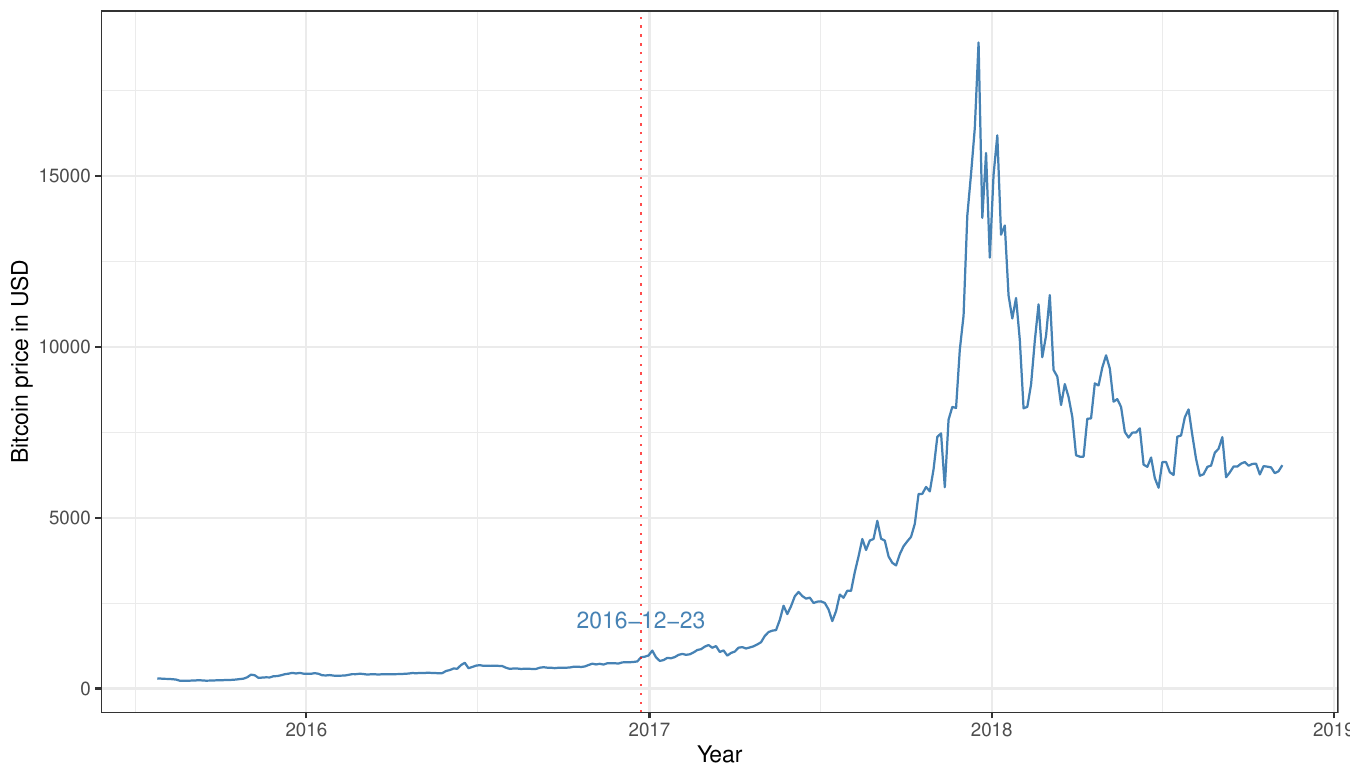}
\caption{Bitcoin market price in USD
  from July 26, 2015, to November 7, 2018, with change point detected on
  December 23, 2016, marked by the red dotted line.}
\label{fig:bitcoin-market-price-original-data}
\end{figure}

We applied the \fct{fastcpd.garch} function to the price data from July 26, 2015, to November 7, 2018. During this period, Bitcoin's price rose from around \$300 to \$19,000 and then fell to around \$6,000. As shown in Figures~\ref{fig:bitcoin-market-price-summary}-\ref{fig:bitcoin-market-price-original-data}, we found a change point at the end of 2016, which is when the Bitcoin price began to rise from around \$700 to \$19,000, more than 25 fold increase.

\subsection{Road casualties in Great Britain from 1969 to 1984}
\label{subsec:road casualties in great britain from 1969 to 1984}

\begin{figure}
\centering
\begin{Schunk}
\begin{Sinput}
R> result_ar <- fastcpd.ar(
+    data = diff(uk_seatbelts[, "drivers"], lag = 12), order = 1, beta = "BIC"
+  )
R> plot(result_ar)
\end{Sinput}
\end{Schunk}
\includegraphics{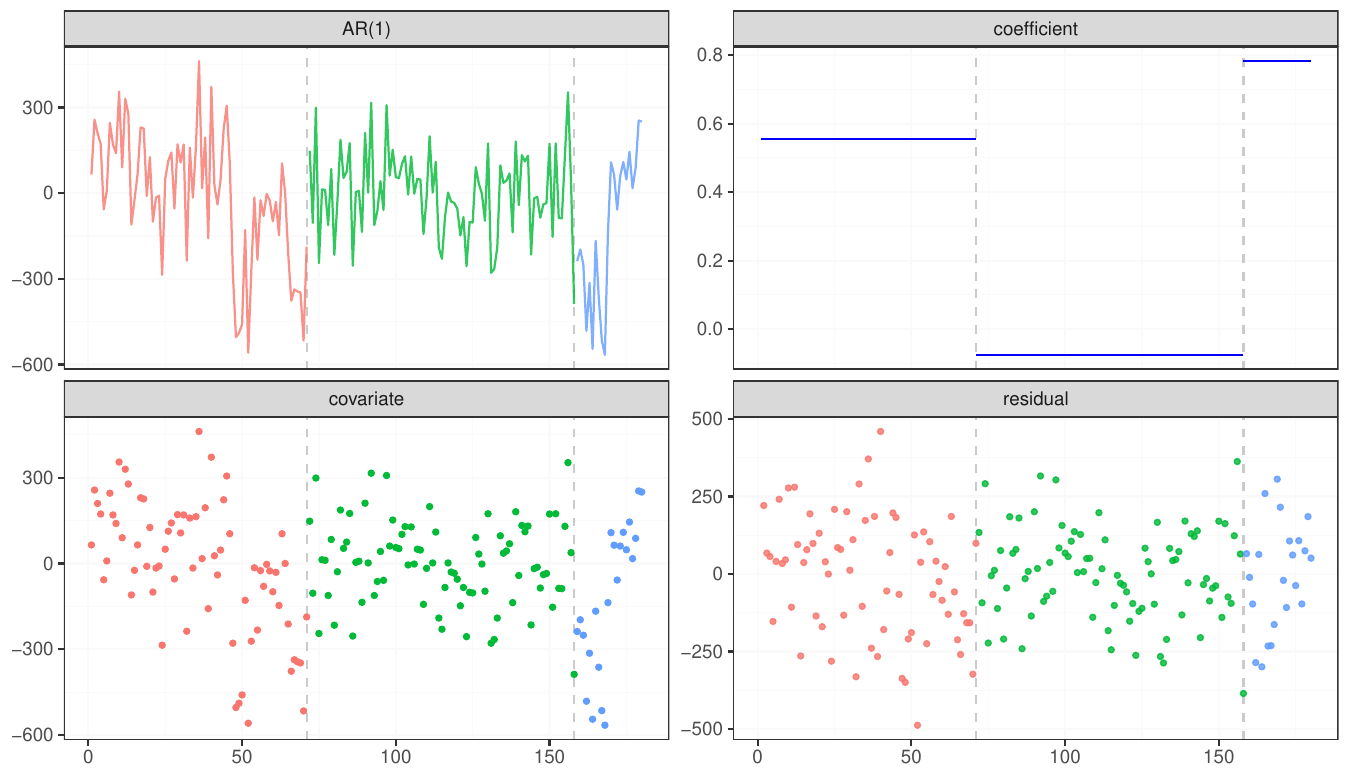}
\caption{Plots of the time series, estimated coefficients, covariate (the covariate here is the same as the data with 1-lag) and residuals together with the estimated change point locations (marked with vertical grey lines) for the UK seatbelts data. The change point analysis is based on a piece-wise AR(1) model for the data after the removal of seasonality.}
\label{fig:road-casualties-in-great-britain-from-1969-to-1984-ar-plot}
\end{figure}

The built-in \pkg{datasets} package in \proglang{R} provides the UK seatbelts data, originally from \cite{harvey1986effects}. This dataset includes records of road casualties in Great Britain from 1969 to 1984. The data contains multiple time series, such as car drivers killed, front-seat and rear-seat passengers killed or seriously injured, distance driven (kms), petrol prices, number of van drivers killed, and a binary indicator for the law's enforcement in a given month. Here, we focus on the \code{drivers} time series and fit a piece-wise AR(1) model to the data \citep{harvey1986effects, zeileis2003testing}.
Figure~\ref{fig:road-casualties-in-great-britain-from-1969-to-1984-ar-plot} summarizes the outputs from the \fct{fastcpd.ar} function with two change points detected, namely April 1974 and November 1982. The two detected change points are possibly related to two associated events: (i) Safety helmets were made compulsory for two-wheeled motor vehicle users from 1973 to 1974; (ii) Seat belt wearing became law for drivers and front seat passengers in 1983; according to Road Casualties Great Britain 2004 Annual Report \citep{statistics2005road}.

Further, we extended our analysis to determine the relationship between car drivers killed and various factors, including distance driven, petrol prices, and the enforcement of seatbelt laws. We consider a linear regression model with piece-wise constant coefficients. The \fct{fastcpd.lm} function detects two change points in April 1974 and November 1982, matching the two change points identified using an AR(1) model.

Inspecting the coefficient changes of ``kms'' (distance driven) before and after April 1974, we can see that the decrease in road casualties can also result from the energy crisis in 1973-1974. During the energy crisis, significant petrol shortages and substantial increases in fuel prices were observed, alongside the implementation of a temporary 50 mph national maximum speed limit \citep{statistics2005road}. These factors likely served as deterrents to unnecessary travel, leading to a reduction in the total kilometers driven by motorists. The combination of higher fuel costs, limited petrol availability, and the imposed speed limit not only contributed to safer driving conditions but also influenced motorists' decisions to undertake shorter or fewer journeys, prioritizing fuel conservation. This shift in driving behavior is corroborated by our analysis, which reveals a significant decrease in the ``kms'' (distance driven) coefficient post-April 1974, aligning with the period's restrictive measures and their impact on road usage patterns.
\begin{Schunk}
\begin{Sinput}
R> result_lm <- fastcpd.lm(
+    diff(uk_seatbelts[, c("drivers", "kms", "PetrolPrice", "law")], lag = 12)
+  )
\end{Sinput}
\end{Schunk}
\begin{figure}
\centering
\includegraphics{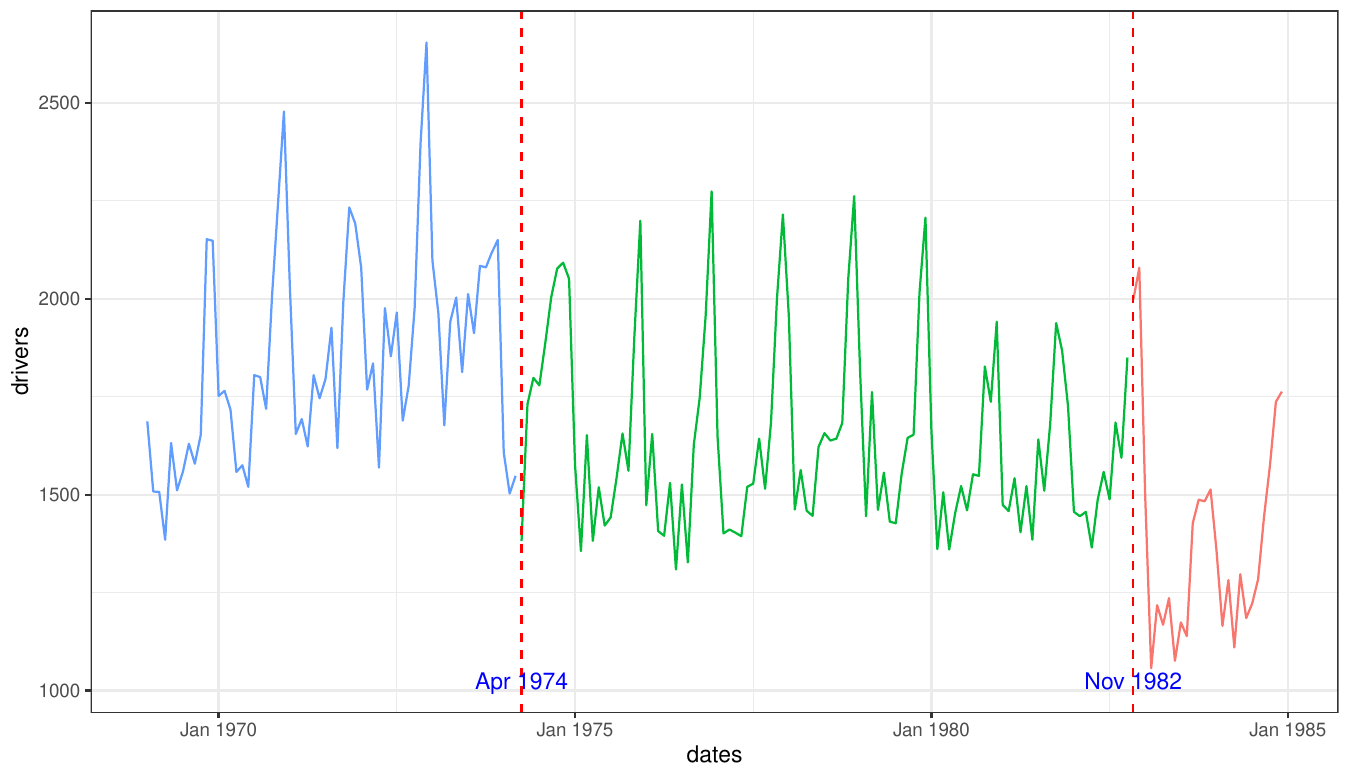}
\caption{Change points detected for the UK seatbelts data using a linear regression model with \code{drivers} as the response and \code{kms}, \code{PetrolPrice} and \code{law} as the predictors.
}
\label{fig:road-casualties-in-great-britain-from-1969-to-1984-lm-plot}
\end{figure}

\section{Advanced usages} \label{sec:advanced usages}

In this section, we will explore some advanced usages of the \pkg{fastcpd} package. Specifically, we will delve into the following topics: (i) interpolating vanilla PELT and SeGD;
(ii) adjusting the number of epochs to fine-tune performance; (iii) adapting line search to mitigate improper gradient update caused by ill-formed Hessian.

\subsection{Interpolating vanilla PELT and SeGD} \label{subsec:advanced usage vanilla pelt}

The \fct{fastcpd} function has a numerical argument called \code{vanilla_percentage}, which is a value (denoted by $v$) between zero and one. For each data segment, if its length is no more than $vT$, the cost value will be computed by performing an exact minimization of the loss function over the parameter. On the other hand, if its length exceeds $vT$, the cost value will be approximated through SeGD. Hence, this parameter creates an algorithm that can be interpreted as an interpolation between dynamic programming with SeGD ($v=0$) and the vanilla PELT ($v=1$). By default, \code{vanilla_percentage} is set to zero. However, if a custom cost function is provided in the form of \code{cost = function(data) \{...\}}, then \code{vanilla_percentage} is set to one, and vanilla PELT is performed.

In practice, it is recommended to set \code{vanilla_percentage} to a small value, which balances the estimation accuracy and computational efficiency. For instance, consider the data generating process in Section~\ref{sec:penalized regression model}, where we set $T = 260$, $p = 40$, and the true change points at $t=65, 130$, and $195$. The other settings remain the same. As we can see from the result, the procedure can better identify the change point locations when setting \code{vanilla_percentage} $=0.5$.
\begin{Schunk}
\begin{Sinput}
R> fastcpd.lasso(small_lasso, segment_count = 2)@cp_set
\end{Sinput}
\begin{Soutput}
[1] 194
\end{Soutput}
\end{Schunk}
\begin{Schunk}
\begin{Sinput}
R> fastcpd.lasso(small_lasso, segment_count = 2, vanilla_percentage = 0.5)@cp_set
\end{Sinput}
\begin{Soutput}
[1]  68 129 193
\end{Soutput}
\end{Schunk}
\begin{figure}
\centering
\includegraphics{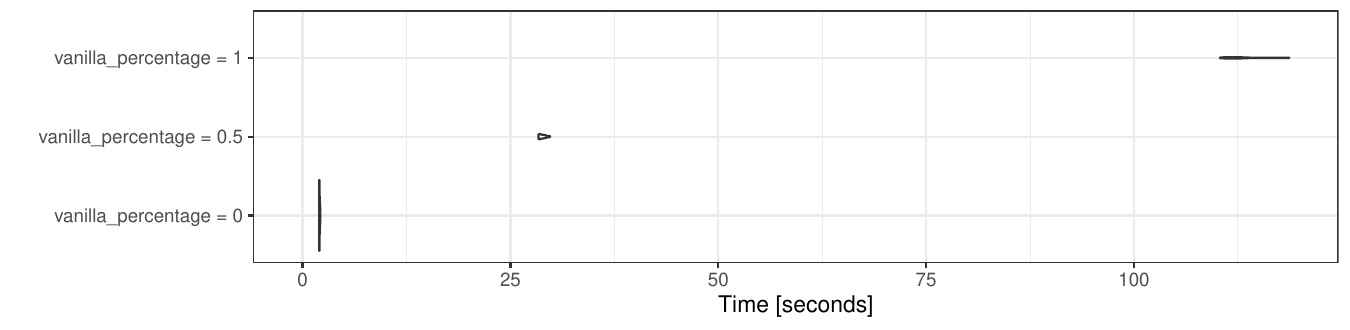}
\caption{Performance comparison between SeGD, interpolated PELT with $v=0.5$ and vanilla PELT.}
\label{fig:vanilla-pelt-0-result-plot}
\end{figure}

\subsection{Adaptive number of epochs}
\label{subsec:advanced usage adaptive number of epochs}

For the SeGD algorithm introduced in Section \ref{sec:algorithm}, a single pass of each data point is performed during the update of the cost value for each segment, which can lead to low data utilization and inaccurate approximation when the length of the data segment is short. Passing the data multiple times (i.e., multiple epochs) will increase data utilization and approximation accuracy to the true cost value. Intuitively, it seems natural to set a larger number of epochs for shorter segments while a smaller number of epochs for longer segments. The idea is to trade the computational speed for better accuracy for shorter segments. As the length of segments increases, the number of epochs gradually decreases to improve the computational efficiency. The \fct{fastcpd} function allows an adaptive scheme by specifying a function argument $W$ to select the number of epochs for each segment. The function $W$ takes the length of the segment as an input and outputs the number of epochs for the segment.
As a concrete example, one can specify the function $W$ as
\begin{equation} \label{eq:adaptive number of epochs}
  W = \left\{ \begin{matrix} 1 & \text{if segment length } \in [0, 100) \\ 0 & \text{if segment length } \in [100, n) \end{matrix} \right.
\end{equation}
with the following code:
\begin{Schunk}
\begin{Sinput}
R> multiple_epochs = function(segment_length) {
+    if (segment_length < 100) 1
+    else 0
+  }
\end{Sinput}
\end{Schunk}
This function will let the SeGD perform parameter updates with an additional epoch for those segments with lengths less than 100. We illustrate the performance of multiple epochs with the function $W$ defined above using the same example as in Section~\ref{subsec:advanced usage vanilla pelt}. Interestingly, the detected change points are closer to the true change points.
\begin{Schunk}
\begin{Sinput}
R> fastcpd.lasso(
+    small_lasso,
+    segment_count = 2,
+    multiple_epochs = function(segment_length) if (segment_length < 100) 1 else 0
+  )@cp_set
\end{Sinput}
\begin{Soutput}
[1]  64 130 195
\end{Soutput}
\end{Schunk}

\subsection{Line search} \label{subsec:advanced usage line search}

The Quasi-Newton method can suffer from potential convergence issues
in scenarios involving ill-formed Hessian matrices.
In such cases, the inversion of the Hessian
matrix in each step of the Quasi-Newton method can result in a large updating step,
disrupting the algorithm's progression and hindering convergence.

To illustrate
the idea, let us revisit the sequential gradient descent step in \eqref{projectedupdatestep}:
\begin{align*}
\hat{\theta}_{\tau+1:t}=\mathcal{P}_{\Theta}(\hat{\theta}_{\tau+1:t-1}-H_{\tau+1:t-1}^{-1}\nabla l(z_t,\hat{\theta}_{\tau+1:t-1})),
\end{align*}
where $-H_{\tau+1:t-1}^{-1}\nabla l(z_t,\hat{\theta}_{\tau+1:t-1})$ is the quasi-Newton's step to update the parameter such that the cost
value estimate is decreasing. An ill-formed Hessian matrix can lead to the unfavorable
result that $\widehat{C}(z_{\tau+1:t}; \hat{\theta}_{\tau+1:t}) > \widehat{C}(z_{\tau+1:t}; \hat{\theta}_{\tau+1:t-1})$.

To tackle this issue, the incorporation of a line search technique becomes
instrumental. Line search is a systematic method employed during optimization
to determine an appropriate step size along the chosen search direction.
In the context of the Quasi-Newton method, line search proves essential for
preventing overly large steps that could otherwise destabilize the optimization
process. By dynamically adjusting the step size based on the local properties
of the objective function, line search enhances the algorithm's robustness and
facilitates convergence in the presence of ill-conditioned Hessian matrices.
Specifically, one can consider the following update:
\begin{align*}
\hat{\theta}_{\tau+1:t}=\mathcal{P}_{\Theta}(\hat{\theta}_{\tau+1:t-1}-\gamma H_{\tau+1:t-1}^{-1}\nabla l(z_t,\hat{\theta}_{\tau+1:t-1})),
\end{align*}
and $\gamma$ is selected such that
\begin{align*}
    \gamma = \argmin_{\gamma \in \Gamma} \{\widehat{C}(z_{\tau+1:t}; \hat{\theta}_{\tau+1:t}) : \hat{\theta}_{\tau+1:t}=\mathcal{P}_{\Theta}(\hat{\theta}_{\tau+1:t-1}-\gamma H_{\tau+1:t-1}^{-1}\nabla l(z_t,\hat{\theta}_{\tau+1:t-1}))\}.
\end{align*}
Here $\Gamma$ is a set that can be specified through the parameter \code{line_search}.
Utilizing line search in gradient descent algorithms can help compare multiple potential updates and find the best updating step to minimize the cost function.


\section{Conclusion} \label{sec:summary}

In conclusion, the \pkg{fastcpd} package stands as a robust and efficient tool for change point detection, offering a versatile framework applicable to a wide spectrum of data types and modeling scenarios. Developed based on the PELT \citep{killick2012optimal} and SeGD \citep{zhang2023sequential}, the algorithm exhibits better computational efficiency, making it well-suited for handling large data sets without compromising estimation accuracy.

In comparison with existing packages, \pkg{fastcpd} outshines in terms of
its built-in functionality, flexibility in specifying the customized cost functions, and computational efficiency. We hope that the \pkg{fastcpd} package will serve as a valuable tool for researchers and practitioners in many real-world applications.


\section*{Computational details}

The results in this paper were obtained using
\proglang{R}~4.3.2 \citep{rlang} with the
\pkg{fastcpd}~0.14.0 package.
\proglang{R} and all packages used are available from the Comprehensive
\proglang{R} Archive Network (CRAN) at
\url{https://CRAN.R-project.org/}. The development version of the package
can be found on GitHub at \url{https://github.com/doccstat/fastcpd}.

\nocite{pierre2015performance}
\nocite{meier2021mosum}

\bibliography{refs}

\newpage

\begin{appendix}

\section{mBIC for general likelihood models}
\label{sec:mbic for mean change with fixed variance}
We derive the mBIC to select the number of change points. We approach this problem by deriving an asymptotic approximation of the Bayes factor. Consider the change point model:
\begin{align*}
z_{i} \sim f(\cdot|\theta_j)\quad  \text{ for } \quad i=\tau_j + 1,\dots,\tau_{j+1},\quad j=0,\dots,k,
\end{align*}
where $\theta_j\in\mathbb{R}^d$, $\tau_0=0$ and $\tau_{k+1}=T.$
Write $Z=(z_1,\dots,z_T)$ and let $\M_k$ be the model with $k$ change points. We aim to find $k$ which maximizes the posterior probability:
\begin{align}\label{eq-1}
\frac{P(\M_k|Z)}{P(\M_0|Z)}=\frac{P(Z|\M_k)P(\M_k)}{P(Z|\M_0)P(\M_0)},
\end{align}
where the first term on the RHS of (\ref{eq-1}) is known as the Bayes factor. We let $\btheta=(\theta_0,\dots,\theta_k)$ and $\btau=(\tau_1,\dots,\tau_k)$ be the parameters associated with model $\M_k$.
We have
\begin{align*}
P(Z|\M_k)=& \int \int P(Z|\btheta,\btau,\M_k)\pi(\btheta,\btau|\M_k) d\btheta d\btau
\\=& \int \int
\exp\left\{
\sum^{k}_{j=0}\sum^{\tau_{j+1}}_{i=\tau_j+1}\log f(z_i|\theta_j)+\log \pi(\btheta|\M_k)\right\}
 d\btheta  \pi(\btau|\M_k) d\btau,
\end{align*}
where $\pi(\btheta,\btau|\M_k)=\pi(\btheta|\M_k)\pi(\btau|\M_k)$ is a prior distribution over the parameters given $\M_k$. We assume that $\btheta\in\Theta_k$ and $\pi(\btheta|\M_k)=C_k$ is the uniform distribution over $\Theta_k$, where $\Theta_k$ is a compact parameter space.
Let $\hat{\theta}_j=\argmax_{\theta} \sum^{\tau_{j+1}}_{i=\tau_j+1}\log f(z_i|\theta).$
By the Taylor expansion, we obtain for each $j$
\begin{align*}
\sum^{\tau_{j+1}}_{i=\tau_j+1}\log f(z_i|\theta_j)
\approx \sum^{\tau_{j+1}}_{i=\tau_j+1}\log f(z_i|\hat{\theta}_j)
-\frac{1}{2}(\theta_j-\hat{\theta}_j)^\top H_j(\hat{\theta}_j)(\theta_j-\hat{\theta}_j),
\end{align*}
where $H_j(\theta)=-\sum^{\tau_{j+1}}_{i=\tau_{j}+1}\partial^2 \log f(z_i|\theta)/\partial \theta\partial \theta^\top$.
Using the Laplace approximation, we get
\begin{align*}
&\int
\exp\left\{
\sum^{k}_{j=0}\sum^{\tau_{j+1}}_{i=\tau_j+1}\log f(z_i|\theta_j)+\log \pi(\btheta|\M_k)\right\}
 d\btheta
 \\=&C_k
\prod^{k}_{j=0}\int \exp\left\{-\frac{1}{2}(\theta_j-\hat{\theta}_j)^\top H_j(\hat{\theta}_j)(\theta_j-\hat{\theta}_j)\right\} d\theta_j \prod^{\tau_{j+1}}_{i=\tau_j+1}f(z_i|\hat{\theta}_j)
 \\=&C_k (2\pi)^{(k+1)d/2}
\prod^{k}_{j=0}|H_j(\hat{\theta}_j)|^{-1/2} \prod^{\tau_{j+1}}_{i=\tau_j+1}f(z_i|\hat{\theta}_j).
\end{align*}
Setting $n_j=\tau_{j+1}-\tau_j$, we have
\begin{align*}
\frac{1}{n_j}H_j(\hat{\theta}_j) \approx I_j(\hat{\theta}_{j}),
\end{align*}
where $I_j(\theta)$ is the Fisher information matrix.
It thus implies that
\begin{align*}
P(Z|\M_k)\approx & C_k (2\pi)^{(k+1)d/2} \int \prod^{k}_{j=0} n_j^{-d/2}|I_j(\hat{\theta}_j)|^{-1/2} \prod^{\tau_{j+1}}_{i=\tau_j+1}f(z_i|\hat{\theta}_j) \pi(\btau|\M_k) d\btau
\\=&C_k (2\pi)^{(k+1)d/2} \int \exp\{U_k(\btau)\} \pi(\btau|\M_k) d\btau,
\end{align*}
where we have defined
\begin{align*}
&\log\left\{\prod^{k}_{j=0} n_j^{-d/2}|I_j(\hat{\theta}_j)|^{-1/2} \prod^{\tau_{j+1}}_{i=\tau_j+1}f(z_i|\hat{\theta}_j)\right\}
\\=&-\frac{d}{2}\sum^{k}_{j=0}\log n_j-\frac{1}{2}\sum^{k}_{j=0}\log(|I_j(\hat{\theta}_j)|)+\sum^{k}_{j=0}\sum^{\tau_{j+1}}_{i=\tau_j+1}\log f(z_i|\hat{\theta}_j)
\\ \approx &-\frac{d}{2}\sum^{k}_{j=0}\log n_j+\sum^{k}_{j=0}\sum^{\tau_{j+1}}_{i=\tau_j+1}\log f(z_i|\hat{\theta}_j):=U_k(\btau).
\end{align*}
The last approximation is due to the assumption that $\log(|I_j(\hat{\theta}_j)|)=o(n_j)$. Next, we discuss the prior on the change point locations. Consider the set
\begin{align*}
\mathcal{D}_k=\left\{\btau: 1< \tau_1<\tau_2<\cdots<\tau_k< T, \min_{0\leq i\leq k}|\tau_{i+1}-\tau_i|\geq \epsilon T\right\},
\end{align*}
for some $\epsilon>0.$ We assume a uniform prior over $\mathcal{D}_k$.
Then,
\begin{align*}
\pi(\btau|\M_k) \approx O\left(\frac{a_{T,k}}{T^k}\right)
\end{align*}
where $\log(a_{T,k}/T^k)=-k\log(T)(1+o(1))$. Let $\hat{\btau}=\argmax_{\btau} U_k(\btau)$. We shall approximate $\int \exp\{U_k(\btau)\} \pi(\btau|\M_k) d\btau$ by
$a_{T,k}\exp\{U_k(\hat{\btau})\}/T^k$, which leads to
\begin{align*}
\log P(Z|\M_k)\approx -k\log(T) + U_k(\hat{\btau}).
\end{align*}
We remark that this approximation can be justified rigorously in the univariate Gaussian model with changing mean and constant variance, see, e.g., \cite{zhang2007modified}.
Combining the above derivations, we suggest choosing $k$ such that
\begin{align*}
-2\sum^{k}_{j=0}\sum^{\tau_{j+1}}_{i=\tau_j+1}\log f(z_i|\hat{\theta}_j) + d\sum^{k}_{j=0}\log n_j +2 k\log(T)
\end{align*}
is minimized, which is equivalent to minimizing the mBIC in Definition \ref{def:mbic}.

\subsection{mBIC for Gaussian location models}

We demonstrate that the proposed mBIC in
\eqref{definition1eq} coincides with the results
in \cite{zhang2007modified} for the mean change model with Gaussian noise.

Suppose an univariate ($d = 1$) data sequence $\{z_t\}_{t = 1}^T$ satisfies
\begin{align*}
  z_t = \mu_t + \epsilon_t,\quad \epsilon_t \sim \mathcal{N}(0, 1),\quad 1 \le t \le T,
\end{align*}
where the mean $\{\mu_t\}_{t = 1}^T$ is piece-wise constant with $k$ change points
$0 = \tau_0 < \tau_1 < \cdots < \tau_k < \tau_{k + 1} = T$. Thus, we have
\begin{align*}
  \mathrm{mBIC}(k,\boldsymbol{\tau}) :=& \sum^{k}_{j=0}C(z_{\tau_{j}+1:\tau_{j+1}}) + \frac{d}{2}\sum^{k}_{j=0}\log (\tau_{j+1}-\tau_j) + (k+1)\log(T) \\
  =& -\sum_{j = 0}^k \sum_{i = \tau_j + 1}^{\tau_{j + 1}} \log f(z_i \vert \hat{\mu}_{\tau_j+1:\tau_{j+1}}) + \frac{1}{2}\sum^{k}_{j=0}\log n_j + (k+1)\log(T) \\
  =& \frac{1}{2}\sum_{j = 0}^k \sum_{i = \tau_j + 1}^{\tau_{j + 1}} (z_i - \hat{\mu}_{\tau_j+1:\tau_{j+1}})^2 + \frac{1}{2}\sum^{k}_{j=0}\log n_j + (k+1)\log(T) + \frac{T}{2}\log(2\pi) \\
  =& \frac{1}{2}\sum_{i = 1}^T (z_i - \hat{\mu}_{\tau_j+1:\tau_{j+1}})^2 + \frac{1}{2}\sum^{k}_{j=0}\log n_j + (k+1)\log(T) + \frac{T}{2}\log(2\pi)\\
  =& \frac{1}{2}\left\{\sum_{i = 1}^T (z_i - \hat{\mu}_{\tau_j+1:\tau_{j+1}})^2 - \sum_{i = 1}^T z_i^2 + T\bar{z}^2\right\} + \frac{1}{2}\sum^{k}_{j=0}\log n_j + \left(k-\frac{1}{2}\right)\log(T) + c_T \\
  =& -\frac{1}{2}\sum_{i = 1}^T (\bar{z} - \hat{\mu}_{\tau_j+1:\tau_{j+1}})^2 + \frac{1}{2}\sum^{k}_{j=0}\log n_j + \left(k-\frac{1}{2}\right)\log(T) + c_T \\
  =& -\frac{1}{2}\sum_{j = 0}^k n_j(\hat{\mu}_{\tau_j+1:\tau_{j+1}} - \bar{z})^2 + \frac{1}{2}\sum^{k}_{j=0}\log n_j + \left(k-\frac{1}{2}\right)\log(T) + c_T,
\end{align*}
where
\begin{align*}
    \bar{z} &= \frac{1}{T} \sum_{t = 1}^T z_i, \\
    n_j &= \tau_{j + 1} - \tau_j, \\
    \hat{\mu}_{\tau_j+1:\tau_{j+1}} &= \frac{1}{n_j}\sum_{t = \tau_j + 1}^{\tau_{j + 1}}z_t,\quad \text{for } \tau_j+1\le i \le \tau_{j+1},
\end{align*}
and $c_T = \{3\log(T) + T\log(2\pi)+\sum_{i=1}^T z_i^2 - T\bar{z}^2\}/2$ is a constant independent of $k$.

\section{Gradient and Hessian for ARMA models}
\label{sec:gradient and hessian of qmle for arma(p, q) model}
\subsection{ARMA(1, 1) models}
\label{subsec:special case for arma(1, 1) model}
Under the assumption that $\epsilon_0 = 0$ and $x_0 = 0$, we have
\begin{align*}
  \epsilon_1 &= x_1, \\
  \epsilon_2 &= x_2 - \phi x_1 - \psi \epsilon_1
              = x_2 - \phi x_1 - \psi x_1, \\
  \epsilon_3 &= x_3 - \phi x_2 - \psi \epsilon_2
              = x_3 - \phi x_2 - \psi (x_2 - \phi x_1 - \psi x_1), \\
  \epsilon_t &= x_t - \phi x_{t - 1} - \psi \epsilon_{t - 1}.
\end{align*}
Denote the noise terms as
\begin{align*}
  \boldsymbol{\epsilon}(\phi, \psi) &=
                  (\epsilon_1, \epsilon_2, \ldots, \epsilon_T)^\top \\
  &= (x_1, x_2 - \phi x_1 - \psi \epsilon_1, x_3 - \phi x_2 - \psi \epsilon_2, \ldots,
      x_T - \phi x_{T - 1} - \psi \epsilon_{T - 1})^\top.
\end{align*}
We can approximate the negative log-likelihood function by the following
Quasi Maximum Likelihood Estimation (QMLE) method:
\begin{align*}
  \tilde{\ell}(\phi, \psi) = \frac{T}{2} \left\{
    \log(2 \pi) + \log(\sigma^2) \right\} +
    \frac{1}{2 \sigma^2}
      \boldsymbol{\epsilon}(\phi, \psi)^\top \boldsymbol{\epsilon}(\phi, \psi).
\end{align*}
The gradient of the negative log-likelihood function with respect to $\phi$,
$\psi$ and $\sigma^2$ are given by
\begin{align*}
  \frac{\partial\tilde{\ell}}{\partial\phi} &= \frac{1}{\sigma^2} \sum_{t = 1}^T c_{\phi}(\psi)_t\epsilon_t,\\
  \frac{\partial\tilde{\ell}}{\partial\psi} &= \frac{1}{\sigma^2} \sum_{t = 1}^T c_{\psi}(\phi, \psi)\epsilon_t,\\
  \frac{\partial\tilde{\ell}}{\partial\sigma^2} &=
    \frac{T}{2 \sigma^2} - \frac{1}{2 \sigma^4} \sum_{t = 1}^T \epsilon_t^2,
\end{align*}
where $c_{\phi}(\psi)_1 = 0$, $c_{\psi}(\phi, \psi)_1 = 0$ and
\begin{align*}
    c_{\phi}(\psi)_t &= \frac{\partial \epsilon_t}{\partial \phi} = - (x_{t - 1} + \psi c_{\phi}(\psi)_{t - 1}), \\
    c_{\psi}(\phi, \psi)_t &= \frac{\partial \epsilon_t}{\partial \psi} = -(\epsilon_{t-1}+\psi c_{\psi}(\phi, \psi)_{t-1}),
\end{align*}
are defined recursively.
The Hessian matrix of the negative log-likelihood function with respect to
$\phi$, $\psi$ and $\sigma^2$ are given by
\begin{align}
  \begin{split}
  \frac{\partial^2\tilde{\ell}}{\partial\phi^2} &= \frac{1}{\sigma^2} \sum_{t = 1}^T c_{\phi}(\psi)_t^2, \\
  \frac{\partial^2\tilde{\ell}}{\partial\phi\partial\psi} &= \frac{1}{\sigma^2} \sum_{t = 1}^T c_{\phi}(\psi)_t c_{\psi}(\phi, \psi)_t + \frac{1}{\sigma^2} \sum_{t = 1}^T c_{\phi, \psi}(\psi)_t \epsilon_t, \\
  \frac{\partial^2\tilde{\ell}}{\partial\sigma^2\partial\phi} &= -\frac{1}{\sigma^4} \sum_{t = 1}^T c_{\phi}(\psi)_t \epsilon_t, \\
  \frac{\partial^2\tilde{\ell}}{\partial\psi^2} &= \frac{1}{\sigma^2}\sum_{t = 1}^T c_{\psi}(\phi, \psi)_t^2 + \frac{1}{\sigma^2} c_{\psi, \psi}(\phi, \psi) \epsilon_t, \\
  \frac{\partial^2\tilde{\ell}}{\partial\sigma^2\partial\psi} &= -\frac{1}{\sigma^4} \sum_{t = 1}^T c_{\psi}(\phi, \psi)_t \epsilon_t, \\
  \frac{\partial^2\tilde{\ell}}{{\partial\sigma^2}^2} &= -\frac{T}{2\sigma^4} + \frac{1}{\sigma^6} \sum_{t = 1}^T \epsilon_t^2,
  \end{split} \label{arma11modelhessianexpression}
\end{align}
where $c_{\phi, \psi}(\psi)_1 = 0$, $c_{\psi, \psi}(\phi, \psi)_1 = 0$ and
\begin{align*}
    c_{\phi, \psi}(\psi)_t &= \frac{\partial^2 \epsilon_t}{\partial \phi \partial \psi} = -(c_{\phi}(\psi)_{t - 1} + \psi c_{\phi, \psi}(\psi)_{t - 1}), \\
    c_{\psi, \psi}(\phi, \psi)_t &= \frac{\partial^2 \epsilon_t}{\partial \psi^2} = - (2 c_{\psi}(\phi, \psi)_{t-1} + \psi c_{\psi, \psi}(\phi, \psi)_{t - 1}),
\end{align*}
are defined recursively.

\subsection{ARMA(\texorpdfstring{$p$}{p}, \texorpdfstring{$q$}{q}) models}
\label{subsec:general case for arma(p, q) model}

For a general ARMA($p$, $q$) model, under the assumption that $\epsilon_{0:1-q} = \boldsymbol{0}$ and $x_{0:1-p} = \boldsymbol{0}$, we have
\begin{align*}
  \epsilon_1 &= x_1, \\
  \epsilon_2 &= x_2 - \boldsymbol{\phi}^\top x_{2-1:2-p} - \boldsymbol{\psi}^\top \epsilon_{2-1:2-q}, \\
  \epsilon_t &= x_t - \boldsymbol{\phi}^\top x_{t - 1:t - p} - \boldsymbol{\psi}^\top \epsilon_{t - 1:t - q},
\end{align*}
and the quasi-negative log-likelihood function can be written as
\begin{align*}
  \tilde{\ell}(\boldsymbol{\phi}, \boldsymbol{\psi}) = \frac{T}{2}
  \left \lbrace
    \log(2 \pi) + \log(\sigma^2) \right \rbrace +
    \frac{1}{2 \sigma^2}
      \boldsymbol{\epsilon}(\boldsymbol{\phi}, \boldsymbol{\psi})^\top
      \boldsymbol{\epsilon}(\boldsymbol{\phi}, \boldsymbol{\psi}),
\end{align*}
where
\begin{align*}
  \boldsymbol{\phi} &= (\phi_1, \phi_2, \ldots, \phi_p)^\top,\\
  \boldsymbol{\psi} &= (\psi_1, \psi_2, \ldots, \psi_q)^\top,\\
  \boldsymbol{\epsilon}(\boldsymbol{\phi}, \boldsymbol{\psi}) &=
    (\epsilon_1, \epsilon_2, \ldots,
      \epsilon_T)^\top \\
  &= (x_1, x_{2} -
      \boldsymbol{\phi}^\top \boldsymbol{x}_{2-1:2-p} -
      \boldsymbol{\psi}^\top \boldsymbol{\epsilon}_{2-1:2-q}, \ldots, x_T - \boldsymbol{\phi}^\top \boldsymbol{x}_{T - 1:T - p} -
      \boldsymbol{\psi}^\top \boldsymbol{\epsilon}_{T - 1:T-q})^\top.
\end{align*}
Thus, the gradient of the negative log-likelihood function with respect to
$\boldsymbol{\phi}$, $\boldsymbol{\psi}$ and $\sigma^2$ are given by
\begin{alignat*}{4}
  \frac{\partial\tilde{\ell}}{\partial\boldsymbol{\phi}} &= \frac{1}{\sigma^2} \sum_{t = 1}^T \epsilon_t \boldsymbol{c}_{\boldsymbol{\phi}}(\boldsymbol{\psi})_t \in \mathbb{R}^{p \times 1}, \\
  \frac{\partial\tilde{\ell}}{\partial\boldsymbol{\psi}} &= \frac{1}{\sigma^2} \sum_{t = 1}^T \epsilon_t \boldsymbol{c}_{\psi}(\boldsymbol{\phi}, \boldsymbol{\psi})_t \in \mathbb{R}^{q \times 1}, \\
  \frac{\partial\tilde{\ell}}{\partial\sigma^2} &= \frac{T}{2 \sigma^2} -
    \frac{1}{2 \sigma^4} \sum_{t = 1}^T \epsilon_t^2 \in \mathbb{R},
\end{alignat*}
where $\boldsymbol{c}_{\boldsymbol{\phi}}(\boldsymbol{\psi})_i = \boldsymbol{0}_p$, for $i = 1-p, \ldots, 1$, $\boldsymbol{c}_{\boldsymbol{\psi}}(\boldsymbol{\phi}, \boldsymbol{\psi})_j = \boldsymbol{0}_q$ for $j=1-q, \ldots, 1$ and
\begin{alignat*}{4}
    \boldsymbol{c}_{\boldsymbol{\phi}}(\boldsymbol{\psi})_t &= \frac{\partial \epsilon_t}{\partial \boldsymbol{\phi}} = -(x_{t-1:t-p} + \sum_{j = 1}^q \psi_j \boldsymbol{c}_{\boldsymbol{\phi}}(\boldsymbol{\psi})_{t-j}) \in \mathbb{R}^{p \times 1}, \\
    \boldsymbol{c}_{\boldsymbol{\psi}}(\boldsymbol{\phi}, \boldsymbol{\psi})_t &= \frac{\partial \epsilon_t}{\partial \boldsymbol{\psi}} = - (\epsilon_{t-1:t-q} + \sum_{j = 1}^q \psi_j \boldsymbol{c}_{\boldsymbol{\psi}}(\boldsymbol{\phi}, \boldsymbol{\psi})_{t-j}) \in \mathbb{R}^{q \times 1},
\end{alignat*}
are defined recursively.
The Hessian matrix of the negative log-likelihood function with respect to
$\boldsymbol{\phi}$, $\boldsymbol{\psi}$ and $\sigma^2$ are given by
\begin{alignat*}{4}
  \frac{\partial^2\tilde{\ell}}{
    \partial\boldsymbol{\phi}^2
  } &= \frac{1}{\sigma^2} \sum_{t = 1}^T \boldsymbol{c}_{\boldsymbol{\phi}}(\boldsymbol{\psi}) \boldsymbol{c}_{\boldsymbol{\phi}}(\boldsymbol{\psi})^\top \in \mathbb{R}^{p \times p},\\
  \frac{\partial^2\tilde{\ell}}{
    \partial\boldsymbol{\phi}\partial\boldsymbol{\psi}
  } &= \frac{1}{\sigma^2} \sum_{t = 1}^T \boldsymbol{c}_{\boldsymbol{\psi}}(\boldsymbol{\phi}, \boldsymbol{\psi})_t \boldsymbol{c}_{\boldsymbol{\phi}}(\boldsymbol{\psi})_t^\top + \frac{1}{\sigma^2} \sum_{t = 1}^T \epsilon_t \boldsymbol{c}_{\boldsymbol{\phi}, \boldsymbol{\psi}}(\boldsymbol{\psi})_t \in \mathbb{R}^{q \times p},\\
  \frac{\partial^2\tilde{\ell}}{
    \partial\sigma^2\partial\boldsymbol{\phi}
  } &= -\frac{1}{\sigma^4} \sum_{t = 1}^T \epsilon_t \boldsymbol{c}_{\boldsymbol{\psi}}(\boldsymbol{\psi})_t \in \mathbb{R}^{p \times 1},\\
  \frac{\partial^2\tilde{\ell}}{
    \partial\boldsymbol{\psi}^2
  } &= \frac{1}{\sigma^2} \sum_{t = 1}^T \boldsymbol{c}_{\boldsymbol{\psi}}(\boldsymbol{\phi}, \boldsymbol{\psi})_t \boldsymbol{c}_{\boldsymbol{\psi}}(\boldsymbol{\phi}, \boldsymbol{\psi})_t^\top + \frac{1}{\sigma^2} \sum_{t = 1}^T \epsilon_t \boldsymbol{c}_{\boldsymbol{\psi}, \boldsymbol{\psi}}(\boldsymbol{\phi}, \boldsymbol{\psi})_t \in \mathbb{R}^{q \times q},\\
  \frac{\partial^2\tilde{\ell}}{
    \partial\sigma^2\partial\boldsymbol{\psi}
  } &= -\frac{1}{\sigma^4} \sum_{t = 1}^T \epsilon_t \boldsymbol{c}_{\boldsymbol{\psi}}(\boldsymbol{\phi}, \boldsymbol{\psi})_t \in \mathbb{R}^{q \times 1},\\
  \frac{\partial^2\tilde{\ell}}{
    {\partial\sigma^2}^2
  } &= -\frac{T}{2 \sigma^4} + \frac{1}{\sigma^6} \sum_{t = 1}^T \epsilon_t^2 \in \mathbb{R},
\end{alignat*}
where $\boldsymbol{c}_{\boldsymbol{\phi}, \boldsymbol{\psi}}(\phi)_i = \boldsymbol{0}_{q\times p}$ for $i = 1-p, \ldots, 1$, $\boldsymbol{c}_{\boldsymbol{\psi}, \boldsymbol{\psi}}(\boldsymbol{\phi}, \boldsymbol{\psi})_j = \boldsymbol{0}_{q \times q}$ for $j = 1-q,\ldots, 1$ and
\begin{alignat*}{4}
    \boldsymbol{c}_{\boldsymbol{\phi}, \boldsymbol{\psi}}(\boldsymbol{\psi})_t &= \frac{\partial^2 \epsilon_t}{\partial \boldsymbol{\phi} \partial \boldsymbol{\psi}} = -(\boldsymbol{c}_{\boldsymbol{\phi}}(\boldsymbol{\psi})_{t-1:t-q} + \sum_{j = 1}^q \psi_j \boldsymbol{c}_{\boldsymbol{\phi}, \boldsymbol{\psi}}(\boldsymbol{\psi})_{t-j})\in \mathbb{R}^{q\times p},\\
    \boldsymbol{c}_{\boldsymbol{\psi}, \boldsymbol{\psi}}(\boldsymbol{\phi}, \boldsymbol{\psi}) &= \frac{\partial^2 \epsilon_t}{\partial \boldsymbol{\psi}^2} = -(\boldsymbol{c}_{\boldsymbol{\psi}}(\boldsymbol{\phi}, \boldsymbol{\psi})_{t-1:t-q} + \boldsymbol{c}_{\boldsymbol{\psi}}(\boldsymbol{\phi}, \boldsymbol{\psi})_{t-1:t-q}^\top + \sum_{j = 1}^q \psi_j \boldsymbol{c}_{\boldsymbol{\psi}, \boldsymbol{\psi}}(\boldsymbol{\phi}, \boldsymbol{\psi})_{t-j}) &&\in \mathbb{R}^{q\times q}.
\end{alignat*}

\end{appendix}

\newpage

\end{document}